\documentclass{emulateapj}
\usepackage{epstopdf}
\usepackage{rotating}
\RequirePackage[running]{lineno}

\shortauthors{Mann, Gaidos, \& Gaudi}
\shorttitle{The Invisible Majority}

\newcommand{\firstw}{-.2cm}
\newcommand{\secondw}{-.3cm}
\newcommand{\thirdw}{-.3cm}
\newcommand{\fourthw}{-.3cm}
\newcommand{\fifthw}{-.3cm}
\newcommand{\sixthw}{-.1cm}

\begin{document}
\title{The Invisible Majority?  Evolution and Detection of Outer Planetary Systems without Gas Giants}  

\author{Andrew W. Mann\altaffilmark{1}, Eric Gaidos\altaffilmark{2}, B. Scott Gaudi\altaffilmark{3}}

\altaffiltext{1}{Institute for Astronomy, University of Hawai'i, 2680 Woodlawn Dr, Honolulu, HI 96822} 
\altaffiltext{2}{Department of Geology \& Geophysics, University of Hawai'i, 1680 East-West Road, Honolulu, HI 96822} 
\altaffiltext{3}{Department of Astronomy, Ohio State University, 140 W. 18th Ave., Columbus, OH 43120}



\begin{abstract}
We present 230 realizations of a numerical model of planet formation
in systems without gas giants.  These represent a scenario in which
protoplanets grow in a region of a circumstellar disk where water ice
condenses and the surface density of solids is enhanced (the ``ice
line''), but fail to accrete massive gas envelopes before the gaseous
disk is dispersed.  Each simulation consists of a small number of
gravitationally interacting oligarchs (protoplanets) and a much larger
number of small bodies that represent the natal disk of planetesimals.
Time zero of each simulation represents the epoch at which the gas has
disappeared, and the dynamics are integrated for 5 billion years
(Gyr).  We investigate systems with varying initial number of
oligarchs, oligarch spacing, location of the ice line, total mass in
the ice line, and oligarch mean density. Systems become chaotic in $\sim 1$ Myr but settle into stable configurations in 10-100 Myr.  We find: (1) runs consistently produce a 5-9 $M_\Earth$ planet at a semimajor axis of 0.25-0.6 times the position of the ice line, (2) the distribution of planets' orbital eccentricities is distinct from, and skewed toward lower values than the observed distribution of (giant) exoplanet orbits, (3) inner systems of two dominant planets (e.g., Earth and Venus) are not stable or do not form because of the gravitational
influence of the innermost icy planet.  The planets predicted by
our model are unlikely to be detected by current Doppler observations.  Microlensing is currently sensitive to the most massive planets found in our simulations, and may have already found several analogs.  A scenario where up to $60\%$ of stars host systems such as those we simulate is consistent with all the available data.  We predict that, if this scenario holds, the NASA {\it Kepler} spacecraft will detect about 120 planets by two or more
transits over the course of its 3.5~yr mission.  Furthermore, we
predict detectable transit timing variations exceeding 20~min due to the presence of additional outer planets.   Future microlensing surveys will detect $\sim 130$ analogs over a $5$~yr survey, including a handful of multiple-planet systems.
Finally, the Space Interferometry Mission (SIM-Lite) should be capable
of detecting $96\%$ of the innermost icy planets over the course of a 5~yr mission.
\end{abstract}
\keywords{celestial mechanics --- planets and satellites: dynamical
evolution and stability --- planets and satellites: formation ---
planet-disk interactions --- planetary systems}

\section{Introduction}
	
The overwhelming majority of more than 400 exoplanets detected
thus far are gas giants with masses comparable to Saturn or Jupiter
\citep{2008PASP..120..531C}.  Most methods of planet
detection, including the Doppler radial velocity
technique which has detected the overwhelming majority of the known exoplanets, are biased toward higher planet mass as well as shorter orbital period \citep{1998ApJ...500..940N, 2003A&A...407..369U, 2004MNRAS.354.1165C}.  Doppler surveys are seriously incomplete for semimajor axes larger than 5~AU, but an extrapolation of a debiased Doppler sample with a ``flat'' distribution predicts that only $\sim$17$\%$ of planetary systems contain giant planets within 20~AU \citep{2003ApJ...598.1350L, 2008PASP..120..531C}.  By analyzing the sample of microlensing planet detections in a survey of high-magnification events, \cite{2010arXiv1001.0572G} estimate that $36\% \pm 15\%$ of the host stars (with typical mass of $\sim 0.5~M_\odot$) host giant planets, with $0.02~M_J$ $< M < 5~M_J$ per logarithmic decade in separation and mass.  By combining the results of \cite{2008PASP..120..531C} for planets with $a<2$~AU with those of \citet{2010arXiv1001.0572G} for $2~$AU $ \la a \la 20~$AU, we estimate that $37\% \pm 13\%$ of stars host giant planets ($m \ga 0.3~M_J$).  We conclude that the fraction of stars hosting giant planets with $a<20$~AU is likely to be at least $20\%$ but less than $50\%$.  

Thus a substantial fraction, and probably the majority of stars do not host giant planets within $\la 20~{\rm AU}$.  However, studies of star-forming regions of different ages have shown that all or nearly all solar-mass stars begin their lives with disks \citep{2001ApJ...553L.153H,  2006AJ....131.1574L, 2006ApJ...638..897S, 2008ApJ...675.1375L, 2010ApJS..186..111L,2010A&A...516A..52M}.  If the accretion of solids is sufficiently rapid and efficient in these disks, then planetary systems are presumably equally numerous around middle-aged stars and the relative paucity of gas giants demands explication.  This conclusion is also supported by the limited statistics of debris disks around solar mass stars \citep{2008ARA&A..46..339W,2009ApJS..181..197C}.

A simple explanation is that gas giants never form around the majority of stars.  The core accretion theory of giant planet formation predicts this outcome if disk gas usually disperses before the growth
of a sufficiently massive solid core triggers runaway accretion of the
gas.  The threshold mass is currently thought to be at least
5~$M_{\oplus}$ \citep{2005Icar..179..415H} but depends sensitively on
the gas opacity \citep{2009MNRAS.393...49A}.  Canonical models of
accretion in a minimum-mass Solar Nebula (MMSN) fail to produce a
sufficiently massive core at the orbit of Jupiter \citep{1996Icar..124...62P} in the $\sim$2-6~Myr
timescale on which disks are observed to dissipate
\citep{2001ApJ...553L.153H, 2005Icar..179..415H, 2009ApJS..181..321E}.  A common explanation for the
formation of Jupiter is to include an ``ice
line'' at 3-5~AU where water condenses and
the surface density of solids is substantially elevated, promoting
core growth \citep{1988Icar...75..146S,1996Icar..124...62P,2002ApJ...581..666K,2004ApJ...616..567I}.  Disks around more massive and/or
more metal-rich stars presumably have greater amounts of solids,
accelerating core accretion \citep{2004ApJ...612L..73L,2009ApJ...694L.171C}, consistent with the
observed correlation of giant planet frequency with host star mass
and metallicity \citep{1999MNRAS.308..447G, 2005ApJ...622.1102F, 2007ApJ...670..833J,2008ApJ...673..502K}.  Theoretical studies also predict a correlation between disk surface density and the frequency (and mass) of giant planets and lower mass icy planets \citep{2004AJ....128.1348R,2009ApJ...690L.140K, 2010ApJS..188..242K}.

A second explanation is that giant planet formation may be stymied by
the tendency of cores to migrate inwards where there is insufficient
gas to form a giant planet \citep{2008ApJ...685..584I}.  This scenario
is predicated on the operation of Type I migration in which torques
from the gas disk become important for bodies more massive than Mars.  The
planet-metallicity correlation is explained if cores in disks with
more solids grow more rapidly to the threshold of runaway gas
accretion, thereby opening up a gap in the disk and halting Type I (but not Type II) migration.
However, the magnitude and even the sign of Type I migration remains very
uncertain \citep{2009ApJ...690L..52L, 2009ApJ...701...18M,2010ApJ...712..198Y}

Another possibility is that giant planets are ubiquitous but have
migrated or been scattered outward
\citep{2009ApJ...705L.148C,2009ApJ...696.1600V, 2009ApJ...693L.113S} to distances
where they are detectable only by microlensing when they behave like isolated lenses \citep{1999ApJ...512..564D,1999ApJ...512..579D, 2007arXiv0704.0767G}, or by their infrared emission under exceptional circumstances of youth, mass, and proximity to Earth
\citep{2006A&A...451..351D,2008Sci...322.1345K,2008Sci...322.1348M,
2009ApJ...707L.123T}.  Finally, giant planets may migrate inwards to
disruption within the Roche zone \citep{2002ApJ...568L.117P}, although there are limits on the ubiquity of such occurrences \citep{2001ApJ...556L..59P,2002AJ....124..400Q}.

The first two explanations predict that systems lacking gas giants
will contain ``failed'' cores of Earth to Neptune mass that
preferentially formed near the ice line \citep{2004ApJ...604..388I, 2008Sci...321..814T, 2008ApJ...673..502K, 2009A&A...501.1161M}. Unless such objects migrated
inward, they would remain invisible to the Doppler technique: the
signal from a 10~$M_{\oplus}$ planet at 5~AU is 0.6~m~s$^{-1}$, well
below the stability of radial velocity measurements on decadal baselines
\citep{2008PASP..120..531C}.  However, the gravitational
microlensing technique is capable of uncovering such planets and several have already been found.  Indeed, \cite{2010ApJ...710.1641S} argue that the
slope of the mass (ratio) function for planets beyond the ice line is
quite steep, such that Neptune-mass planets are $\sim 7$ times more
common than Jupiter-mass planets.  Combined with the \cite{2010arXiv1001.0572G} normalization of the frequency of gas giants in this region, this implies that the majority of stars host Neptune or lower mass
planets beyond the snow line.

In this paper we investigate a scenario in which several protoplanets
or ``oligarchs,'' but not giant planets, have formed at or beyond the
ice line at the time disk gas has disappeared.  A major premise of our
initial conditions is that Type I migration was not effective in these
disks.  (Type II migration does not act on planets much less massive
than Jupiter).  We carry out direct numerical integrations of the
orbital and mass evolution of these protoplanets as they accrete
additional mass from a residual disk of planetesimals.  We integrate
the orbits of the (proto)planets over 5~Gyr to ascertain the stability
of these systems and their configuration at a plausible epoch at which
they might be observed.

Our simulations complement the works of \citet{2008ApJ...673..487I, 2008ApJ...682.1264K, 2009A&A...501.1161M}, and \citet{2008Sci...321..814T}. \cite{2008ApJ...673..487I} and \cite{2009A&A...501.1161M} use analytical models of the orderly growth of cores in disks during the interval that gas is
present.  They include migration due to torques exerted by the disk,
but neglect subsequent, chaotic gravitational interactions between the
cores and the residual disk of solids \citep{2008ApJ...685..584I}.  \cite{2008Sci...321..814T} model
the dynamical interactions between protoplanets/cores as well as the
gas disk but analytically proscribe accretion from a fixed disk of
solids.  In cases of low disk mass
or rapid gas removal, all three investigations predict formation of Earth- to Neptune-sized bodies. Like \citet{2008ApJ...682.1264K}, our simulations {\it begin} at the end of the orderly growth
phase when the gas has disappeared, and {\it assume} that no giant
planets have formed, but unlike them, we assume that a massive residual disk of smaller bodies is still present.

Our work also contrasts with investigations of the dynamical evolution
and configuration of systems of {\it giant} planets like those detected by Doppler surveys,
e.g., \citet{1996Sci...274..954R}, \citet{2003Icar..163..290A}, \citet{2008ApJ...686..580C}, \citet{2009ApJ...699L..88R}, and \cite{2010ApJ...711..772R}.  We expect that the evolution of systems of solid
planets that emerge from the icy part of a disk will differ
substantially, primarily because the Safronov number
\begin{equation}
S = \frac{v_{\mbox{\footnotesize esc}}^2}{2v_{\mbox{\footnotesize orb}}^2}
\end{equation}
will be $\le 1$, whereas $S \gg 1$ in systems of giant planets
(excluding ``hot'' Jupiters).  More efficient accretion, less intense
scattering, and stronger coupling to the planetesimal disk are expected.

The goal of our simulations is threefold: first, we want to predict
the evolution and final configuration of such systems for a range of
plausible initial conditions.  Second, we wish to determine if and how
such planets could be detected by present or future means, i.e., the
{\it Kepler} and Space Interferometry Missions (SIM-Lite) and ground-based
microlensing surveys.  Lastly, we want to establish how measurements
of these objects might be used to infer the initial conditions and histories of
these systems, especially important given how poorly planet formation
is understood.

\section{Methods and Models}\label{sec:methodsmodels}

\subsection{Approach and Assumptions \label{sec:assumedscenario}}

\begin{figure}[t] 
   \centering
   \includegraphics[width=8.5cm]{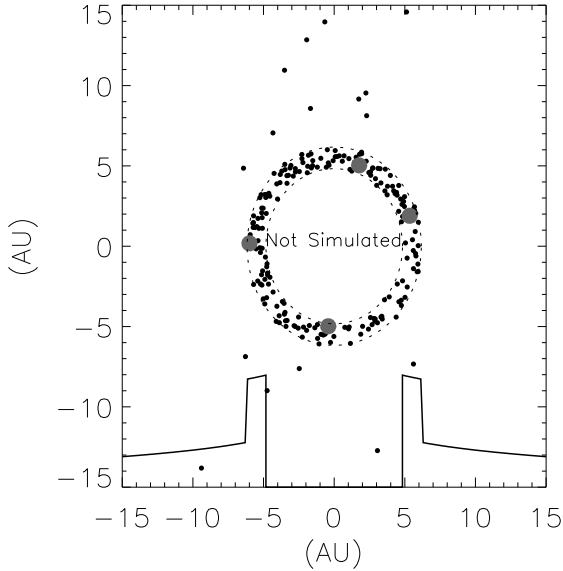} 
   \caption{Schematic of the typical initial configuration of our systems.  Oligarchs are shown in gray in a zone of enhanced surface density between 5 and 6 AU.  Small bodies are shown in black and represent a much larger number of planetesimals in the disk.  Plotted along the bottom of the graph is an approximation of the logarithm of the surface density of material.  Beyond the region of oligarch formation, the density of planetesimals follows a $\Sigma \propto a^{-1}$ mass distribution.  The inner system is left empty because this region has only a minor effect on the evolution of the outer system, but requires considerably more time to simulate.}
   \label{fig:assumescenario}
\end{figure}

Each numerical realization consists of a $1M_{\sun}$ central star
surrounded by a planet-forming disk (Figure \ref{fig:assumescenario}).
Our scenario assumes the canonical theory of planet formation which
consists of three phases: (1) runaway accretion of protoplanets from a
disk of planetesimals; (2) slower oligarchic growth of these
protoplanets as they consume neighboring planetesimals and
each other; and (3) a chaotic or giant impact phase when the
mass in residual planetesimals falls below that in the protoplanets
and the oligarchs' orbits begin to cross \citep{2004ApJ...614..497G,
2006AJ....131.1837K}.  The disk includes a region of width $\delta_{\mbox{\footnotesize ice}}$
immediately beyond the ice line ($a_{\mbox{\footnotesize ice}}$) in which the surface
density of solids is enhanced by the transport of water vapor out of
the inner disk and condensation as ice at $a > a_{\mbox{\footnotesize ice}}$ \citep{2004ApJ...614..490C, 2006Icar..181..178C}.  We assume
that oligarchic protoplanets have appeared in this region because of
the enhanced density and the relatively short orbital time scale
compared to the disk further out.  A background disk of unincorporated
planetesimals extends from $a_{\mbox{\footnotesize ice}}$ to 400~AU following the surface
density profile $\Sigma \propto a^{-1}$ found in observations of
protostars \citep{2007ApJ...659..705A}.  For computational efficiency,
we exclude the region outside of 100~AU and the region inside the ice
line.  A few sets of simulations included inner bodies to assay their
effect on the formation of planets further out (Section
\ref{sec.validation}).

We assume a disk of solar composition with the total mass within 400~AU of $0.03$~$M_{\sun}$ and a corresponding mass of condensible solids (rock and ice) of $\sim$ $150~M_{\Earth}$ \citep{2003ApJ...591.1220L}.  The mass of the
background disk within the 100~AU simulation region is
43-48~$M_\earth$, depending on how much mass is moved into the ice
line.  The additional mass added to the ice line is varied (Section
\ref{sec.values}).  We specify the number of oligarchs $n$ and the
spacing between them in Hill radii $b$; this sets the mass of each
oligarch and the total mass in oligarchs.  The remaining mass, in fact
the majority of the mass in all simulations, is distributed evenly
among small bodies that represent primordial planetesimals.  The
number of small bodies is limited by computational resources and is
500 in all of our simulations, except for a single run with 1000.  The
mass of each small body is about $0.08M_\earth$, much less than the
oligarch masses.  We do not model the fragmentation of planetesimals,
the production of dust by a collisional cascade, and the
removal of that dust by stellar radiation \citep{2008ARA&A..46..339W}.  We discuss the possible
consequences of fragmentation on our conclusions in Section \ref{sec:limitations}.
Oligarchs and small bodies were given non-zero
random inclinations ($|i| < 10^{-3}$ and $10^{-2}$ degrees,
respectively) and eccentricities ($e \leq 10^{-3}$ and $10^{-2}$,
respectively), although we experiment with higher initial $i$ and $e$ values in a single simulation set (see Section \ref{sec.validation}).

\begin{figure}[t] 
   \centering
   \includegraphics[width=8.5cm]{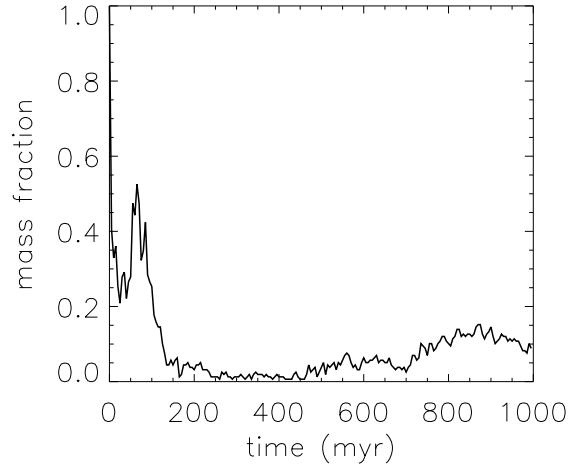} 
   \caption{Fraction of mass in small bodies within the orbit of the outermost oligarch during the first 1 Gyr for a run from the ST3 set.  More than $95\%$ of the mass is in the small bodies at the start of the simulation.  By 500 Myr the region has been almost completely cleared of small bodies.  Only as the outer oligarch migrates outward do more small bodies enter the region.}
   \label{fig:massden}
\end{figure}

The equations of motion of each particle were integrated using the hybrid integrator code Mercury6 with the combination of a second-order, mixed-variable symplectic integrator and the Burlirsch--Stoer integrator for close encounters \citep{1999MNRAS.304..793C}.  
Mercury6 divides the simulation into large and small bodies.  Large bodies interact gravitationally with, and can collide with both large and small bodies.  Small bodies also interact gravitationally and collide with large bodies, but they cannot collide with other small bodies.
Computations were performed on the
TeraGrid network \citep{teragrid:2007}.  Integrations were performed
for 5~Gyr, except for more computationally intensive runs used to
check our conditions (Section \ref{sec.validation}).  Ten replicate
simulations were performed for each set of parameters.  The initial
positions and orbits were varied slightly between simulations in a
set.  A timestep of 40 days was used for most simulations, based on the expected
orbital period of the innermost oligarch and the requirement that
there be at least 10-20 time steps per orbital period, a conservative
setting \citep{2010ApJ...711..772R}.  Simulations that included oligarchs in the inner system had a timestep of 8 days, and simulations with a close-in ice line were given a timestep of 20 days.  Runs where there were fewer than 10
timesteps per oligarch orbit for an extended period of time were either adjusted, rerun, or had short
test simulations run parallel to them to test the accuracy of the results.

\begin{figure}[htbp] 
   \centering
   \includegraphics[width=8.5cm]{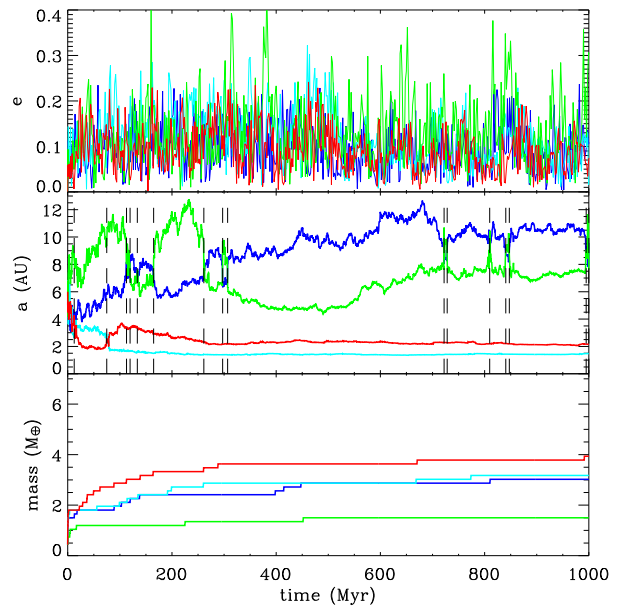} 
   \caption{Evolution of the system in a run from the ST4 set.  Oligarch swaps (exchange of order with
distance from the star) are marked with dashed lines, excluding the first 10 Myr when oligarch swaps are more frequent.  Because oligarchs often undergo numerous swaps when they approach each other's Hill radius, each dashed line may represent multiple swaps.}
   \label{fig:evo1}
\end{figure}

\begin{figure}[htbp] 
   \centering
   \includegraphics[width=8.5cm]{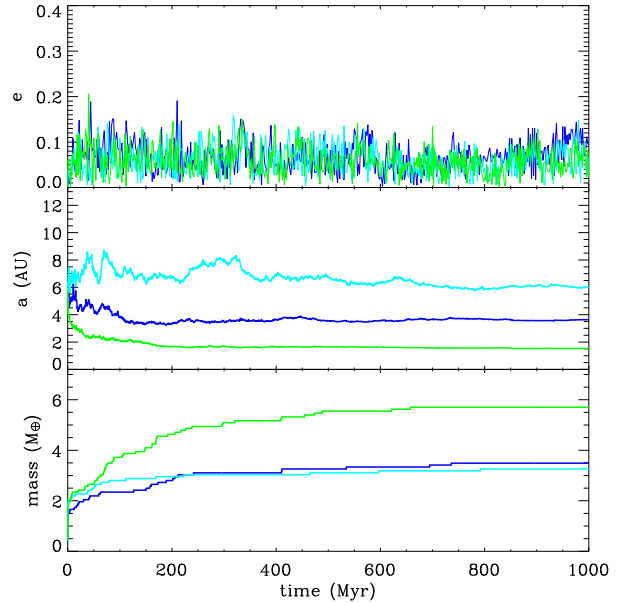} 
   \caption{Evolution of the system in a run from the ST3 set showing the migration of the innermost migratory planet (IMP) (green).  Initially, the IMP (green) migrate inward, while the second and third oligarch (blue and teal respectively) migrate outward through their exchange of angular momentum.  All three oligarchs are driven inward at times by their interactions with the planetesimal disk.  After $\sim 100$~Myr the region from 0 to 6 AU has been almost completely cleared of small bodies, causing the inner two oligarchs to settle into relatively stable orbits.  This type of angular momentum exchange is common in simulations that start with 3-4 oligarchs.}
   \label{fig:evo2}
\end{figure}

For computational efficiency, small bodies in most runs are removed after 1~Gyr.  Previous work has shown that this economy will not significantly affect the evolution of the oligarchs if the mass
surface density of the small bodies is much less than that of the
oligarchs \citep{2006AJ....131.1837K}.  Figure \ref{fig:massden} shows
the fraction of mass in small bodies inside the orbit of the outermost
oligarch for the first Gyr of a run in our standard star.  Although
at the start of the simulation the small bodies represent the bulk of
the mass, by 100~Myr they are less than $50\%$ of the mass in the
oligarch region.  By 1~Gyr, small bodies account for only $\sim 10\%$
of the total mass.  We do not observe significant changes in the
orbits of the oligarchs as a result of the removal of the
small bodies at 1 Gyr.  Oligarchs tend to move into stable orbits long before 1 Gyr.  This can be seen in Figures \ref{fig:evo1} and
\ref{fig:evo2}.  In a single run from each set
we retain the small bodies for 2 Gyr as a check.

\subsection{Parameter Values and Initial Conditions \label{sec.values}}

\begin{table}[htdp]
\begin{scriptsize}
\begin{center}
\caption{Simulation Initial Conditions}
\label{tab:initialstats}
\begin{tabular}{ l c c c c c l }
\hline
	\hline
    Name & \hspace{\firstw}$b$  &\hspace{\secondw}$n$ & \hspace{\thirdw}$M_{\mbox{\footnotesize ice}}$ & \hspace{\fourthw}$a_{\mbox{\footnotesize ice}}$ & \hspace{\fifthw}$m_{i}$  &\hspace{\sixthw}`$Comments$ \\ 
    &$\hspace{\firstw}(R_H)$&\hspace{\secondw}&$\hspace{\thirdw}(M_{\earth})$&\hspace{\fourthw}(AU) &\hspace{\fifthw}$(M_{\earth})$&\hspace{\sixthw} \\
    	\tableline
    ST2&\hspace{\firstw}8&\hspace{\secondw}2&\hspace{\thirdw}35&\hspace{\fourthw}5.0&\hspace{\fifthw}0.44&\hspace{\sixthw}Standard \\
    ST3&\hspace{\firstw}8&\hspace{\secondw}3&\hspace{\thirdw}35&\hspace{\fourthw}5.0&\hspace{\fifthw}0.44&\hspace{\sixthw}Standard \\
    ST4&\hspace{\firstw}8&\hspace{\secondw}4&\hspace{\thirdw}35&\hspace{\fourthw}5.0&\hspace{\fifthw}0.44&\hspace{\sixthw}Standard \\
	\hline
    LM2&\hspace{\firstw}8&\hspace{\secondw}2&\hspace{\thirdw}10&\hspace{\fourthw}5.0&\hspace{\fifthw}0.44&\hspace{\sixthw}Low ice line mass\\
    LM3&\hspace{\firstw}8&\hspace{\secondw}3&\hspace{\thirdw}10&\hspace{\fourthw}5.0&\hspace{\fifthw}0.44&\hspace{\sixthw}Low ice line mass\\
    LM4&\hspace{\firstw}8&\hspace{\secondw}4&\hspace{\thirdw}10&\hspace{\fourthw}5.0&\hspace{\fifthw}0.44&\hspace{\sixthw}Low ice line mass\\
	\hline
    CI2&\hspace{\firstw}8&\hspace{\secondw}2&\hspace{\thirdw}35&\hspace{\fourthw}2.7&\hspace{\fifthw}0.44&\hspace{\sixthw}Close ice line\\
    CI3&\hspace{\firstw}8&\hspace{\secondw}3&\hspace{\thirdw}35&\hspace{\fourthw}2.7&\hspace{\fifthw}0.44&\hspace{\sixthw}Close ice line\\
    CI4&\hspace{\firstw}8&\hspace{\secondw}4&\hspace{\thirdw}35&\hspace{\fourthw}2.7&\hspace{\fifthw}0.44&\hspace{\sixthw}Close ice line\\
	\hline
    MP2&\hspace{\firstw}5&\hspace{\secondw}2&\hspace{\thirdw}35&\hspace{\fourthw}5.0&\hspace{\fifthw}1.78&\hspace{\sixthw}``Medium packed'' oligarchy\\
    MP3&\hspace{\firstw}5&\hspace{\secondw}3&\hspace{\thirdw}35&\hspace{\fourthw}5.0&\hspace{\fifthw}1.78&\hspace{\sixthw}``Medium packed'' oligarchy\\
    MP4&\hspace{\firstw}5&\hspace{\secondw}4&\hspace{\thirdw}35&\hspace{\fourthw}5.0&\hspace{\fifthw}1.78&\hspace{\sixthw}``Medium packed'' oligarchy\\	
	\hline
    HP7&\hspace{\firstw}2&\hspace{\secondw}7&\hspace{\thirdw}35&\hspace{\fourthw}5.0&\hspace{\fifthw}0.70&\hspace{\sixthw}``Highly packed'' oligarchy\\
    HP9&\hspace{\firstw}2&\hspace{\secondw} 9&\hspace{\thirdw}35&\hspace{\fourthw}5.0&\hspace{\fifthw}1.98&\hspace{\sixthw}``Highly packed'' oligarchy\\  
    HP12&\hspace{\firstw}2&\hspace{\secondw}12&\hspace{\thirdw}35&\hspace{\fourthw}5.0&\hspace{\fifthw}3.63&\hspace{\sixthw}``Highly packed'' oligarchy\\
	\hline
    LD2&\hspace{\firstw}8&\hspace{\secondw}2&\hspace{\thirdw}35&\hspace{\fourthw}5.0&\hspace{\fifthw}0.44&\hspace{\sixthw} Low density oligarchs\\
    LD3&\hspace{\firstw}8&\hspace{\secondw}3&\hspace{\thirdw}35&\hspace{\fourthw}5.0&\hspace{\fifthw}0.44&\hspace{\sixthw} Low density oligarchs\\
    LD4&\hspace{\firstw}8&\hspace{\secondw}4&\hspace{\thirdw}35&\hspace{\fourthw}5.0&\hspace{\fifthw}0.44&\hspace{\sixthw} Low density oligarchs\\
    \hline
    RD3\footnotemark[1]&\hspace{\firstw}$\sim8$&\hspace{\secondw}3&\hspace{\thirdw}10&\hspace{\fourthw}5&\hspace{\fifthw}0.30-0.70&\hspace{\sixthw}Random $m$, $a$ of oligarchs\\
    TS3&\hspace{\firstw}8&\hspace{\secondw}3&\hspace{\thirdw}35&\hspace{\fourthw}5&\hspace{\fifthw}0.44&\hspace{\sixthw}1000 small bodies\\
    EV3\footnotemark[2]&\hspace{\firstw}8&\hspace{\secondw}3&\hspace{\thirdw}35&\hspace{\fourthw}5&\hspace{\fifthw}0.44&\hspace{\sixthw}Earth, Venus included\\
    IO3\footnotemark[3]&\hspace{\firstw}8&\hspace{\secondw}3&\hspace{\thirdw}35&\hspace{\fourthw}5&\hspace{\fifthw}0.44&\hspace{\sixthw}Oligarchs in inner system\\
    HE3\footnotemark[4]&\hspace{\firstw}8&\hspace{\secondw}3&\hspace{\thirdw}35&\hspace{\fourthw}5&\hspace{\fifthw}0.44&\hspace{\sixthw}Higher initial $e$, $i$\\
    	\hline
\end{tabular}
\begin{minipage}{8.5cm}
	\footnotemark[1]{The mass of the oligarchs in this system varies randomly by $10\%$ and the semimajor axis by $0.2~AU$.  $b$ was allowed to vary based on the mass and semimajor axis, but it was not allowed to go below $b=6$ or above $b=9$.}    \\
	\footnotemark[2]{This simulation set was run to 100 Myr with a planetesimal disk and then to 1 Gyr without small bodies.}\\
	\footnotemark[3]{This system has 130 oligarchs in the inner system following a $b=8$ and $\Sigma \propto a^{-1}$ distribution.  The total mass of the inner system was $2.2~M_\Earth$ spread between $0.5$ and $5~AU$.  This simulation set was run to 100 Myr.} \\
	\footnotemark[4]{This system has small bodies with $|i| < 10^{-1}$ and $e \leq 5\times10^{-2}$ for the small bodies in the region of the oligarchs.  This simulation set was run to 250 Myr.}
	\end{minipage}
\end{center}
\end{scriptsize}
\end{table}%

Our simulations are described by four principal parameters: $a_{\mbox{\footnotesize ice}}$, $M_{\mbox{\footnotesize ice}}$, $b$, and $n$.
We also vary the mass density of the oligarchs $\rho$.  Table
\ref{tab:initialstats} lists the parameters for our 18 sets, each of which consists of 10 replicate runs.  Five additional sets of 10 sensitivity simulations are shown at the bottom of the table and discussed in Section \ref{sec.validation}.

{\it Ice line location:} The location of the ice line $a_{\mbox{\footnotesize ice}}$
depends on the opacity and rate of viscous dissipation in the disk and
the mass of the central star, and is time dependent
\citep{2006Icar..181..178C}.  The ice line in the primordial solar
system has been variously placed near 5~AU (to stimulate the rapid
formation of Jupiter's core and explain its icy satellites)
\citep{1988Icar...75..146S} or at 2.7~AU (to coincide with the
transition between hydrated and anhydrous asteroids) \citep{2000orem.book..413A}.  The position of the ice line presumably varies
between planetary disks.  In the majority of our simulations, we set
$a_{\mbox{\footnotesize ice}} = 5$~AU.  In the CI sets (30 runs) $a_{\mbox{\footnotesize ice}}$
was fixed at 2.7~AU.

{\it Ice line mass:} Our choice of mass just beyond the ice line is guided by an estimate of the mass of water transported through the inner disk that re-condensed at the ice line, and the mass of solids in the cores of the outer planets in our solar system.  To the small amount ($< 1$ $M_\earth$) of solids within $a_{\mbox{\footnotesize ice}} < a < a_{\mbox{\footnotesize ice}}+\delta_{\mbox{\footnotesize ice}}$ predicted by a simple $\Sigma \propto a^{-1}$ distribution, we add a fraction of the total amount of water from the disk outside this region.  We adopt a minimum value of $M_{\mbox{\footnotesize ice}}=$ 10~$M_\earth$.   Our maximum value ($M_{\mbox{\footnotesize ice}}=$ 35~$M_\earth$) assumes all four outer solar system planets formed near the ice line \citep{2002AJ....123.2862T} and sums their initial core masses: 10 $M_\earth$ for Jupiter, 15 $M_\earth$ for Saturn \citep{2009sfch.book...75H} and 5 $M_\earth$ (below the critical threshold) each for Uranus and Neptune.  $M_{ice}=$ 10~$M_\earth$ and $35$~$M_\earth$ correspond to $15\%$ and $60\%$ of the disk's water in the ice line and a background disk mass within 100 AU of $43$ and $48~M_{\oplus}$, respectively.  Other estimates of the mass in the ice line are similar \citep{1988Icar...75..146S, 2004A&A...417..151K, 2008ApJ...685..584I}.

{\it Oligarch spacing:} The spacing between oligarchs is specified
as a multiple of the Hill radius,
\begin{equation}
R_H = a\left(\frac{m}{3M_*}\right)^{1/3},
\end{equation}
where $m$ is the initial oligarch mass, and $M_* = 1~M_{\Sun}$.  We
use $b=8$ \citep{2006Icar..180..496C} as well as $b=5$
\citep{ 2007ApJ...661..602F, 2009ApJ...696L..98R,2010ApJ...711..772R}.
Additional runs contain ``overpacked'' oligarchies with
$b=2$ (the approximate Roche limit).  This last value is well within the
instability limit $b=3^{4/3}$ where accretion can be rapid \citep{1993Icar..106..247G} but outside
horseshoe (1:1 resonance) orbits
\citep{2009AJ....137.3778C}.

{\it Numbers and masses of oligarchs:} We vary the number of
oligarchs $n$ between two and four in the $b$ = 5 and $b$ = 8 simulation sets,
in analogy to the number of cores that formed at the ice line in our
solar system.  For the case of $n$ = 3 and $a_{\mbox{\footnotesize ice}} = 5$~AU, we fixed the width of the ice
line $\delta_{\mbox{\footnotesize ice}}$ to 1~AU \citep{1988Icar...75..146S,
2004A&A...417..151K}.  The initial mass of each oligarch is related
to $n$, $a_{\mbox{\footnotesize ice}}$, and $\delta_{\mbox{\footnotesize ice}}$ by

\begin{equation}
\label{eqn.oligarchmass}
m \approx 3M_* \left(\frac{\delta_{ice}}{nba}\right)^3.
\end{equation}  

This gives initial oligarch masses of 0.44 and
1.78~$M_\Earth$ for $b=8$ and $b=5$, respectively.  We subsequently scale
$\delta_{\mbox{\footnotesize ice}}$ with $a_{\mbox{\footnotesize ice}}$ and $n$ so that for a given value of
$b$, the initial oligarch mass is unchanged.  In the overpacked
($b=2$) scenario, the requirement that the core mass not exceed
10~$M_{\earth}$ (and seed giant planet formation) requires $n > 4$.
In these case, we fix the total mass in oligarchs $nm$ as 0.25,
0.50, and 0.75 times the total ice line mass.  We calculate $n$ using
Equation (\ref{eqn.oligarchmass}).  This gives $n=12$, 9 and 7 and
oligarch masses of 0.7, 1.98, and 3.63~$M_\earth$, respectively.  In
any given run, all initial oligarch masses are the same, except
for the RD3 simulation set, in which the oligarch mass is
allowed to vary slightly.

{\it Oligarch mass density:} We use mean densities $\rho$ predicted
by \citet{2009ApJ...693..722G} for planets of $60\%$ ice ($15\%$ for the LM sets) and the rest rock.  However,
super-Earth-mass bodies may retain significant envelopes of H/He gas,
giving them mean densities more akin to that of Neptune ($\rho =
1.64$~g~cm$^{-3}$), and this would increase the  cross-section for accretion.
To investigate this effect, the LD simulation sets were run with
$\rho =1$~g~cm$^{-3}$.

\subsection{Sensitivity Runs \label{sec.validation}}

{\it Larger number of small bodies:} The 500 small bodies in each
simulation represent a much larger population of planetesimals in the
disk.  One run (the TP3 set) includes 1000 small bodies, each
with half the mass of those in the 500-small body runs.  Neither 500 nor 1000 small bodies is physical (there may actually be trillions of planetesimals).  The goal here is to
ascertain if the results depended sensitively on the number of small
bodies or their mass. 

{\it Retaining small bodies for 2 Gyr:} For a randomly selected
run in each set (excluding the CI and
HP sets), the small bodies are kept in the simulation for 2 Gyr
instead of 1 Gyr.  This is done to verify that removal of the small
bodies does not create a bias in the results.

{\it Non-identical oligarch masses:} A real system of protoplanets will not have identically spaced, identical mass oligarchs.  In one set of sensitivity runs (RD3 set) we vary the mass by $\sim$10\% and the semimajor axis by $\pm 0.2$ AU.  In these simulations $b$ is allowed to vary between 7 and 9 as a result of the randomized semimajor axis.

{\it Mass in the inner system:} Two sets of runs investigate the effect of mass in the inner system ($a < a_{\mbox{\footnotesize ice}}$) on planet formation in the outer system.  In one set (IO3), we include a disk of 130 small (0.002-0.1~$M_{\Earth}$) oligarchs with a mass surface density $\Sigma \sim a^{-1}$, separation of $b = 8$, and a total mass of 2.2~$M_{\Earth}$ (the total mass of the inner planets in the solar system).  Given that a common outcome of higher-resolution $N$-body simulations of terrestrial planet accretion are two planets of roughly equal mass \citep{2009Icar..203..644R}, a second set of runs was performed that included two planets with the masses and orbits of Earth and Venus.

{\it Small Body Eccentricities:} Our small bodies start with non-zero but very small values of orbital eccentricity and inclination.  However, gravitational perturbations by the oligarchs will drive these values to finite values ($\sim 0.05$) even in the presence of disk gas \citep{2006Icar..180..496C}.  We run a single set of simulations to 250 Myr with $|i| < 10^{-1}$ and $e \leq 5\times10^{-2}$ for the small bodies in the region of the oligarchs.

\section{Results \label{sec:results}}

\subsection{Dynamical Evolution \label{sec:systemevo}}

We are primarily interested in describing the gross dynamical
evolution of these systems over the first 1 Gyr, especially when small
bodies remain in the system and are being scattered or accreted by the
oligarchs.  At later times, most (but not all) of the systems do not substantially evolve.  We analyze orbital parameters with a resolution
of 100 kyr: important events happen on shorter timescales
than this, especially in the first 1 Myr of the simulations
when systems are most chaotic.  However, analysis of such events is
not the focus of this work.

\begin{table}[htdp]
\begin{scriptsize}
\begin{center}
\caption{Evolution Statistics}
\label{tab:evolutionstats}
\begin{tabular}{ c | r r r r r }
	\hline
		\hline
    Name & $\tau_{clear}$\footnotemark[1] &Resonance& Frac Order&Inner Planet&
    \\
    &(Myr)&Crossings& Preserved& Migration (AU)&\\
    	\hline
    ST2&$290$&100&0.6&3.7& \\
    ST3&$80$&470&0.2&3.8& \\
    ST4&$90$&980&0.2&3.7&\\
	\hline
    LM2&$450$&260&0.7&3.2&\\
    LM3&$290$&820&0.1&3.0&\\
    LM4&$240$&1490&0.1&3.2&\\
	\hline
   CI2&$62$&50&0.4&2.0&\\
    CI3&$65$&320&0.2&1.9&\\
    CI4&$56$&390&0.3&2.0&\\
	\hline
    MP2&$80$&70&0.7&3.8&\\
    MP3&$43$&500&0.6&3.8&\\
    MP4&$36$&1297&0.2&3.8&\\
	\hline
    HP7&$16$&3900&0.1&3.1&\\
    HP9&$26$&11600&0.0&3.3&  \\  
    HP12&$66$&28600&0.0&3.5&\\
	\hline
    LD2&$150$&80&0.6&3.8&\\
    LD3&$74$&600&0.3&3.9&\\
    LD4&$77$&1050&0.3&3.8&\\
    	\hline
\end{tabular}
\begin{minipage}{8.5cm}
\footnotemark[1]{$\tau_{clear}$ is the average time for the disk to lose $70\%$ of it's mass within 10 AU.}    
\end{minipage}
\end{center}
\end{scriptsize}
\end{table}%

\begin{figure}[htbp] 
   \centering
   \includegraphics[width=8.5cm]{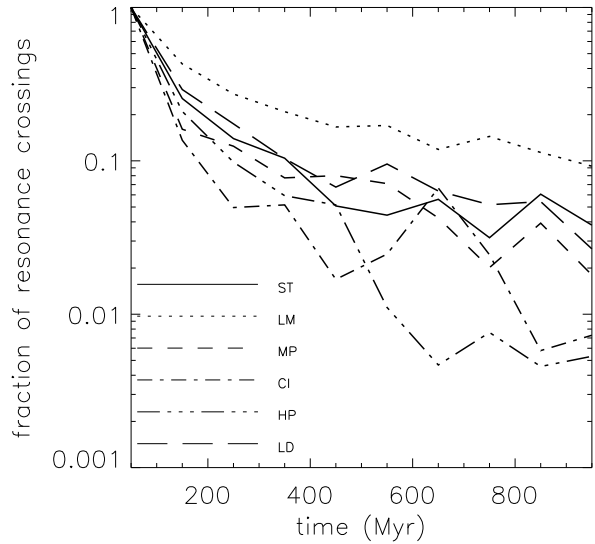} 
   \caption{Number of MMR crossings in 100 Myr bins, normalized by the first bin.  This is a measure of the relative levels of chaos in the system.  For our purposes, a resonance crossing occurs whenever an oligarch crosses a 1:1, 2:1, 3:2, 3:1, 4:1, 5:3, or 5:2 mean motion commensurability with another oligarch.  Total number of resonance crossings are reported in Table \ref{tab:evolutionstats}. }
   \label{fig:resonanceall}
\end{figure}

We use five metrics to describe the evolution of each system (Table
\ref{tab:evolutionstats}): (1) the time $\tau_{clear}$ in which 70\%
of small bodies are removed from inside 10~AU; (2) the number of
mean motion resonance (MMR) crossings experienced by the oligarchs
with other oligarchs over the first 1 Gyr; (3) the fraction of runs
in which two or more oligarchs ``swap'', i.e., exchange order with
distance from the star; (4) the distance of inward migration by the
(ultimately) innermost protoplanet; and (5) the number of
oligarchs ejected from a system.

{\it Disk clearing:} $\tau_{\mbox{\footnotesize clear}}$ is the time over which 70\% of
small bodies within 10~AU are accreted or ejected.  The timescale is
not sensitive to the precise choice of outer boundary.
$\tau_{\mbox{\footnotesize clear}}$ can be as long as several hundred Myr (Table
\ref{tab:evolutionstats}).  $\tau_{\mbox{\footnotesize clear}}$ is shorter in systems with
more oligarchs (more accreting bodies), lower values of $b$ (oligarchs
scatter each other onto more eccentric orbits); a closer ice line
(shorter orbital period and dynamical time scale), and lower oligarch mass
density (greater cross section of accretion).  The LM sets
have longer $\tau_{\mbox{\footnotesize clear}}$ values because there is less
concentration of mass in the ice line, proximal to the oligarchs.  The
oligarchs do not grow as quickly nor scatter as efficiently.

{\it Mean motion resonance crossing:} We count the number of times an
oligarch passes through a 1:1, 2:1, 3:2, 3:1, 4:1, 5:3, or 5:2
mean-motion commensurability with another oligarch (Table
\ref{tab:evolutionstats}).  The greatest number of MMR crossings
occurs in runs with lower values of $b$ and larger $n$.  Figure
\ref{fig:resonanceall} shows the normalized rate of MMR crossings per
time for the six primary simulation groups, binned in 100~Myr
intervals.  In all cases the rate decreases with time and by 1-3
orders of magnitude over the first 1~Gyr as the systems evolve.

{\it Oligarch swapping:} In 43\% of the runs in the ST sets,
at least one swap (where two oligarchs exchange rank in semimajor
axis) occurs between 50~Myr and 1~Gyr.  Swapping occurs when two
oligarchs approach within two Hill radii (the zone of strong
scattering).  The oligarchs can collide, or they can enter horseshoe
orbits (within a single Hill radius) \citep{2009AJ....137.3778C}.  In the latter case, they often exchange places quickly, usually in $\ll 1$ Myr.  Figure \ref{fig:evo1}
shows an extreme case where there are at least 13 distinct swaps over the first
1 Gyr, excluding the chaotic period in the first 50 Myr.

{\it Inner Planet Migration:} In all of our primary runs, an oligarch
migrates inward to a position between 1 and 3 AU, most often settling
between 1.2 and 1.9 AU, i.e. 3-4~AU from its initial starting place.
In $81\%$ of cases this body has grown to become the most massive
planet in the system.  Most runs resemble that of Figure
\ref{fig:evo2}.  Figure \ref{fig:evo1} shows an unusual run in which two
inner oligarchs migrate inward.  The inward migration of one or more
oligarchs is a manifestation of the redistribution of angular
momentum in a circumstellar disk and its resulting radial
dispersal \citep{1981ARA&A..19..137P}.  The angular momentum of the
inwardly migrating oligarch(s) is lost to scattered small bodies and
some of it is transferred to outwardly migrating oligarchs
(Figure \ref{fig:evo2}).  This process will occur as long as there is
a sufficiently massive disk of small bodies, i.e., for tens or hundreds
of Myr (Table \ref{tab:evolutionstats}).  Analogous events may have
unfolded during the early dynamical evolution of the outer Solar
System: Jupiter migrated inward while the other giant planets moved
outward as a result of angular momentum exchange through a residual
disk of planetesimals \citep{1993Natur.365..819M, 1999AJ....117.3041H,
2005Natur.435..466G}.  In our simulations, the Safronov number is less
than or not much greater than one, and significant accretion of mass
occurs during migration.

\citet{2000ApJ...534..428I} formulated the migration rate of a low mass
planet moving through a planetesimal disk as
\begin{equation}
\frac{da}{dt} = \frac{a}{P_K}\frac{4\pi \Sigma_p a^2}{M_*},
\end{equation}
where $P_K$ is the Keplerian orbital period and $M_*$ is the mass of
the central star.  The migration rate is independent of planet mass.  Recast in terms of the ice line mass, the migration timescale is 
\begin{equation}
\tau_{\mbox{\footnotesize migrate}} \approx P_K \frac{2 a}{\Delta}\frac{M_*}{M_{\mbox{\footnotesize ice}}},
\end{equation}
which as short as $\sim 1$~Myr for the undepleted disk.  The observed
migration timescale is slower ($\sim$10~Myr) and
is probably in part due to the depletion of the disk by the oligarchs
themselves, but may also reflect the inability of our simulations with low numbers of particles to correctly resolve the distribution
of planetesimals in horseshoe orbits that most strongly interact with the planet.

\begin{figure}[htb] 
   \centering
   \includegraphics[width=8.5cm]{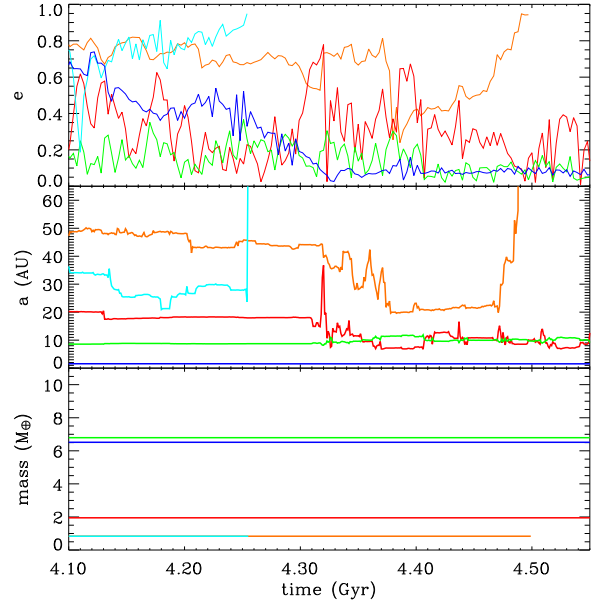} 
   \caption{Evolution of the system in a run from the HP12 set showing the ejection of two oligarchs (teal and orange bodies) at $\sim$ 4.25 Gyr and $\sim$ 4.5 Gyr.}
   \label{fig:evo3}
\end{figure}

{\it Oligarch Ejection}: Ejection of oligarchs can occur during the
final, chaotic phase of planet formation \citep{1987Icar...69..249L, 
1999Natur.400...32S,2007ApJ...668L.167D}.  An oligarch is considered
``ejected'' in our simulations if it attains $a >$ 400~AU.  Ejection of an
oligarch is frequent (13 of 30 runs) in the $b=2$ simulations (Figure
\ref{fig:evo3}).  In most of these cases, an ejection occurs within 1
Myr after two oligarchs appear to enter a resonance.  Resonance
between two oligarchs increases their orbital eccentricities, and also
excites neighboring oligarchs, and this is sometimes
suffcient to eject smaller oligarchs from the system.  In a single
simulation, two oligarchs stayed near resonance for 100~Myr, causing
the ejection of 3 other oligarchs.

\subsection{Configuration at 5~Gyr}\label{sec:outcomes}

\begin{figure*}[htb] 
   \centering
   \includegraphics[height=9.5cm,trim=90 0 0 0, clip=true]{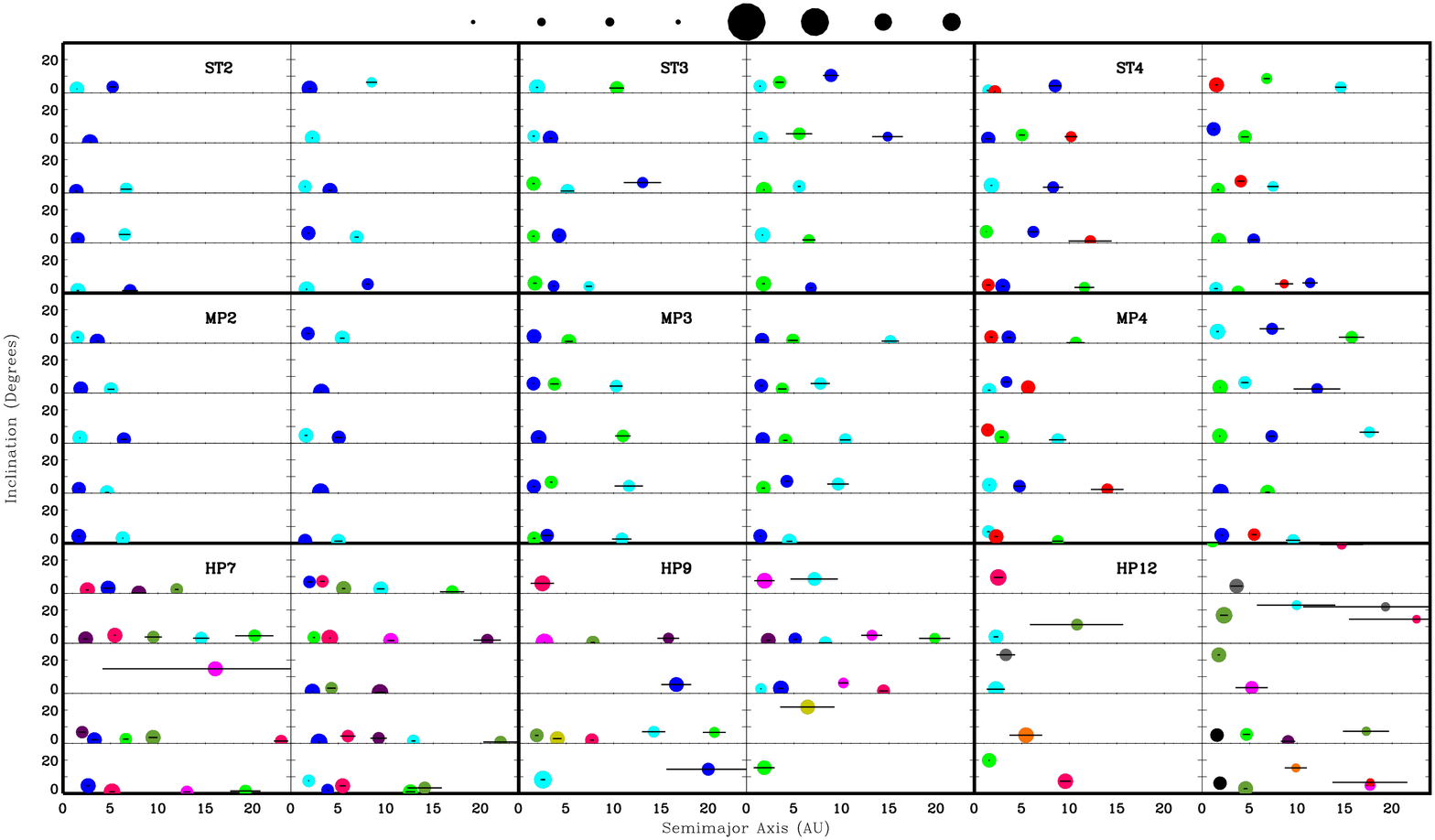} 
      \caption{Final configuration of all simulations in the ST sets (top), MP sets (middle), and HP sets (bottom).  Simulation sets are organized left to right by initial number of oligarchs $n$, with the fewest on the left.  Color coding shows the initial position of the oligarch (blue is initially closest to the star).  Each circle represents a planet scaled in size by the planet mass.  A line going through the point represents the periapsis and apoapsis of its orbit.  As the top is the solar system plotted as a mass (but not distance) scale.  Some simulations from the HP sets had planets outside 25 AU, which cannot be seen in these plots.}	
   \label{fig:finalconfigs}
\end{figure*}

\begin{figure}[htb] 
   \centering
   \includegraphics[width=8.5cm]{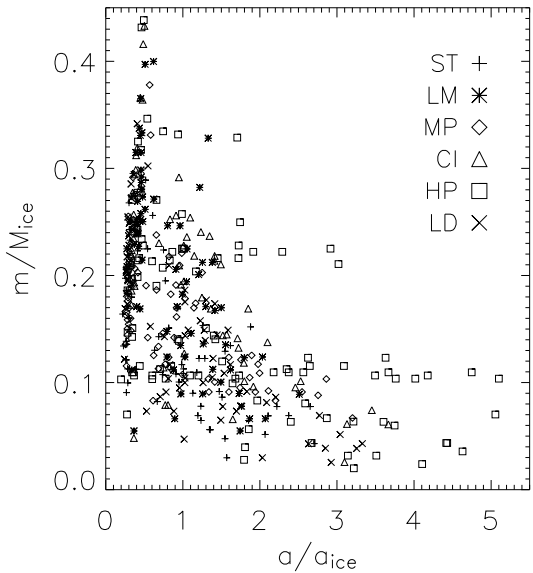} 
   \caption{Plot of mass vs. semimajor axis scaled by the initial mass and position of the ice line, respectively, for the 6 primary simulation groups.  The innermost planets at $0.25$ to $0.60~a_{\mbox{\footnotesize ice}}$ have a clear separation from the other planets.  Simulations show a statistical mass segregation effect out to $\sim 2.5~a_{mbox{\footnotesize ice}}$.}
   \label{fig:masterMA}
\end{figure}

{\it Mass:} Figure \ref{fig:finalconfigs} shows the final system configurations produced
by 9 sets of 10 replicate runs (standard, MP, and HP sets)
after 5~Gyr.  Only 4 of the systems contain a single planet; all
started with only $n=2$ oligarchs.  Mass segregation (the tendency of
higher mass planets to appear closer to the star) occurs under all
conditions (Figure \ref{fig:masterMA}).  Of the 180 primary runs,
$\sim 94\%$ produce a planet that ultimately resides between $ 0.25$ 
and $0.6$ $a_{\mbox{\footnotesize ice}}$, and in 78\% of all $b$ = $8$ runs, this planet is the most
massive one.  This effect is least pronounced in the $b=2$ set.
Approximately one-fourth of the total mass in the ice line is
incorporated into these planets.  In fewer than half of the runs did the
initial innermost oligarch become this innermost planet (Figure
\ref{fig:finalconfigs}), in agreement with \cite{2008ApJ...686..580C},
who found that in systems of equal-mass gas giants, each planet has
roughly equal probability of becoming the innermost.  Planet mass
decreases with semimajor axis between $0.6$ to $2.5~a_{\mbox{\footnotesize ice}}$.  Planets
ending outside $\sim 2.5~a_{\mbox{\footnotesize ice}}$ have masses close to that of the
original oligarch.  These bodies were scattered outside the ice line
early in the run and have accreted little mass.

\begin{figure}[htbp] 
   \centering
   \includegraphics[width=8.5cm]{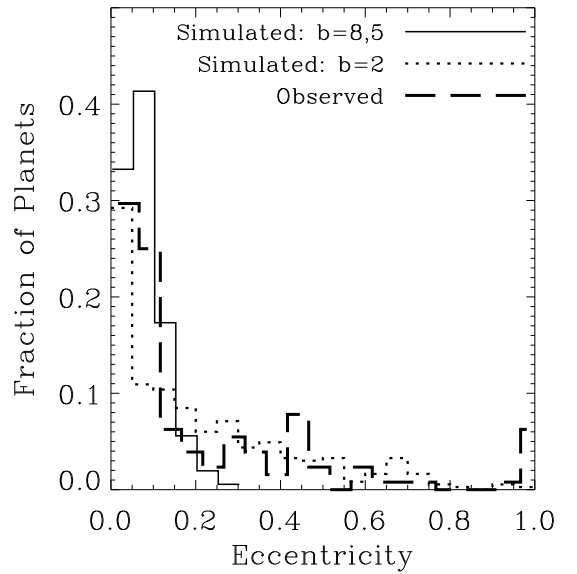} 
   \caption{Distribution of orbital eccentricities of simulated planets for $b=8, 5$, and $b=2$ and known exoplanets (exoplanet.eu).  Only a few of our $b=8, 5$ planets have eccentricities higher than 0.2 whereas the observed gas giant planets span the full range of eccentricities from 0 to nearly 1.  The highly packed ($b=2$) simulations exhibit an eccentricity distribution much closer to observations, which are mostly gas giants.}
   \label{fig:eccentricityhist}
\end{figure}

{\it Eccentricity:} We find no correlation between orbital
eccentricity and mass, semimajor axis, or number of oligarchs in our
runs.  Our runs with $b=2$ (and the most oligarchs) produce a larger
dispersion in eccentricity, but this could be a result of the higher
total mass in oligarchs.  For $b=8$, no planet ended with $e > 0.3$,
and for the ST sets only 2 planets had eccentricities above 0.2.
This contrasts with the observed distribution amongst detected
planets, and simulations of systems of gas giants
\citep{ 2007ARA&A..45..397U,2008ApJ...686..580C, 2008ApJ...686..603J} (Figure
\ref{fig:eccentricityhist}).  Smaller values of eccentricity are
expected in systems where the mass of the oligarchs that perturb each
other is lower relative to the mass of the disk of small bodies that
dampen such perturbations
\citep{2006Icar..180..496C,2008ApJ...687L.107R}.

\begin{table}[htdp]
\begin{center}
\begin{scriptsize}
\caption{Simulation Outcome Classifications}
\label{tab:outcomestats}
\begin{tabular}{c r r r r r r r r }
	\hline
		\hline
    Name & \hspace{\secondw}$\overline{M_{tot}}$ & $\overline{n_f}$ & $\overline{OSS}$ &\hspace{\firstw}$\sigma_{OSS}$ &$\overline{RMC}$&$\hspace{\firstw}\sigma_{RMC}$&$\overline{AMD}$ &$\sigma_{AMD}$ \\ 
    &\hspace{\secondw}($M_\earth$)&&&&&\hspace{\firstw}&&\\
    \hline
ST2& \hspace{\secondw}    10.98&            1.8&10.32&\hspace{\firstw} 1.14&13.76&\hspace{\firstw} 4.75&0.0051&0.0019\\
ST3& \hspace{\secondw}    12.76&            2.4& 8.09&\hspace{\firstw} 1.81&14.10& \hspace{\firstw}5.73&0.0095&0.0070\\
ST4& \hspace{\secondw}    12.12&            2.8& 7.69&\hspace{\firstw} 1.79&10.79&\hspace{\firstw} 4.05&0.0093&0.0036\\
\hline
LM2& \hspace{\secondw}     4.29&            1.8& 9.60&\hspace{\firstw} 1.77&21.44& \hspace{\firstw}3.87&0.0121&0.0076\\
LM3& \hspace{\secondw}     4.93&            2.4& 8.84&\hspace{\firstw} 1.39&17.71& \hspace{\firstw}5.80&0.0117&0.0058\\
LM4&  \hspace{\secondw}    5.08&            2.8& 7.80&\hspace{\firstw} 1.96&18.32& \hspace{\firstw}3.48&0.0152&0.0047\\
\hline
CI2&    \hspace{\secondw} 17.16&            2.0& 8.17&\hspace{\firstw} 1.40&15.61&\hspace{\firstw} 4.97&0.0019&0.0006\\
CI3&   \hspace{\secondw}  16.99&            2.5& 7.32&\hspace{\firstw} 1.48&14.78&\hspace{\firstw} 3.48&0.0051&0.0021\\
CI4&  \hspace{\secondw}   16.39&            2.4& 8.00&\hspace{\firstw} 1.50&13.03&\hspace{\firstw} 4.15&0.0050&0.0031\\
\hline
MP2&  \hspace{\secondw}   12.88&            1.8& 8.41&\hspace{\firstw} 0.98&16.57&\hspace{\firstw} 3.64&0.0038&0.0018\\
MP3& \hspace{\secondw}    14.22&            2.7& 7.43&\hspace{\firstw} 1.60&10.76&\hspace{\firstw} 3.00&0.0080&0.0037\\
MP4& \hspace{\secondw}    15.21&            2.9& 6.67&\hspace{\firstw} 0.99&10.68&\hspace{\firstw} 3.36&0.0076&0.0041\\
\hline
HP7&  \hspace{\secondw}   26.71&            4.6& 3.85&\hspace{\firstw} 0.60& 9.68& \hspace{\firstw}4.00&0.0485&0.1397\\
HP9&  \hspace{\secondw}   18.67&            4.0& 5.86&\hspace{\firstw} 2.19& 5.34& \hspace{\firstw}4.12&0.1560&0.1747\\
HP12& \hspace{\secondw}    14.89&            4.2& 6.11&\hspace{\firstw} 2.32& 4.74& \hspace{\firstw}1.97&0.1413&0.1222\\
\hline
LD2& \hspace{\secondw}    13.45&            1.9&10.12&\hspace{\firstw} 0.70&11.58& \hspace{\firstw}1.58&0.0037&0.0013\\
LD3& \hspace{\secondw}    13.17&            2.6& 8.27&\hspace{\firstw} 1.83&10.47& \hspace{\firstw}2.67&0.0093&0.0049\\
LD4& \hspace{\secondw}    13.55&            3.0& 7.36&\hspace{\firstw} 1.99&10.34& \hspace{\firstw}6.03&0.0093&0.0042\\
	\hline
RD3\hspace{\secondw}&12.01&2.4&8.66&\hspace{\firstw}2.25&14.12&\hspace{\firstw}6.01&0.0030&0.0050\\
TP3&\hspace{\secondw}12.05&            2.4& 7.84&\hspace{\firstw} 1.73&14.78&\hspace{\firstw} 5.34&0.0031&0.0018\\
EV3\footnotemark[1]&\hspace{\secondw}11.23&            3.7& 5.97&\hspace{\firstw} 0.83&10.28&\hspace{\firstw} 1.33&0.0167&0.0096\\
IO3\footnotemark[1]&\hspace{\secondw}11.74&           30.9& 1.11&\hspace{\firstw} 0.11& 8.38&\hspace{\firstw} 4.00&0.0521&0.0124\\
HE3\footnotemark[2]& \hspace{\secondw}     10.55&            2.7& 7.43&\hspace{\firstw} 0.75&21.12&\hspace{\firstw} 6.92&0.0082&0.0038\\
	\hline
\end{tabular}
\begin{minipage}{8.5cm}
	\footnotemark[1]{Configuration at 100~Myr.}    \\
	\footnotemark[2]{Configuration at 200~Myr}    \\
	\end{minipage}	
	\end{scriptsize}
\end{center}
\end{table}%

{\it Dynamical classification: } \cite{2001Icar..152..205C} describes
several dimensionless parameters to compare the outcome of simulations
of accretion in the inner solar system to the actual planets.  We
adopt three, the radial mass concentration RMC, the angular momentum
deficit (AMD), and the orbital spacing statistic (OSS), to classify
and compare our results.  The RMC measures how mass is distributed in
the system and is given by
\begin{equation}
\mbox{RMC} = \mbox{max}\bigg(\frac{\Sigma m_j}{\Sigma m_j.
[\mbox{log}_{10}(a/a_j)]^2}\bigg),
\end{equation}
where $m_j$ and $a_j$ are the masses and semimajor axes of the planets in a system.
A more tightly concentrated system will have a higher RMC.  The AMD is
a measure of orbital excitation and is given by
\begin{equation}
\mbox{AMD} = \frac{\Sigma_j m_j \sqrt{a_j}[1-\sqrt{(1-e_j^2)}\cos i_j]}{\Sigma_j m_j \sqrt{a_j}},
\end{equation}
where $i_j$ and $e_j$ are the inclinations and eccentricities of the planets in a system.
The AMD measures the difference between the angular momentum (in the
z-direction) of a system and that of a system of identical bodies on circular,
non-inclined orbits with the same semimajor axes.  The OSS is a
measure of the mean spacing of planets and is given by
\begin{equation}
\mbox{OSS} =\frac{1}{N-1}\bigg(\frac{a_{\mbox{\footnotesize max}}-a_{\mbox{\footnotesize min}}}{a_{\mbox{\footnotesize max}}+a_{\mbox{\footnotesize min}}}\bigg)\bigg(\frac{3M_*}{2\bar{m}}\bigg)^{1/4},
\end{equation}
where $N$ is the number of bodies, $a_{\mbox{\footnotesize max}}$ is the maximum semimajor
axis, $a_{\mbox{\footnotesize min}}$ is the minimum semimajor axis, $M_*$ is the mass of
the central star, and $\bar{m}$ is the mean mass of the oligarchs.
Unlike the RMC, the OSS ignores the mass and location of individual
oligarchs and depends on the distance between the bodies.  

\begin{figure}[htbp] 
   \centering
   \includegraphics[width=8.5cm]{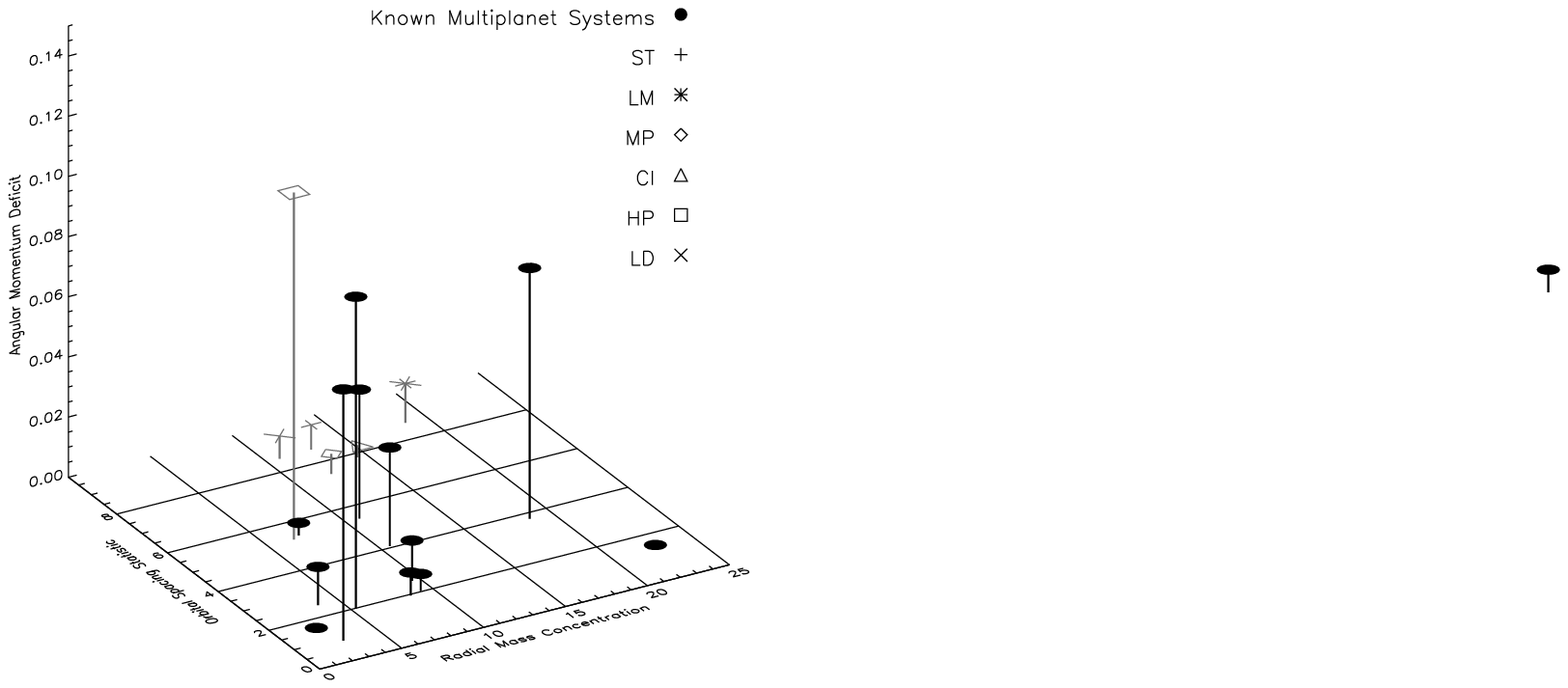} 
   \caption{Average OSS, RMC, and AMD for the 6 primary simulation groups alongside 12 known planetary systems with $\ge 3$ planets as well as the solar system values.  Most of the observed exoplanet systems shown contain at least 1 gas giant.  Since inclinations are measured from an invariable plane, which is not known or poorly defined for exoplanetary systems, we assume zero inclinations for these calculations.  Although the known planets cover a wide range of values, they are clearly very different from our simulations with the exception of the HP ($b=2$) runs.  }
   \label{fig:pca}
\end{figure}

These statistics have no meaning for
single-planet systems and those cases are excluded from the
calculations.  Average values and standard deviations of RMC, AMD, and OSS for each set of simulations are listed in Table \ref{tab:outcomestats} along with values for a number of exoplanetary systems and the solar system.  These data are also plotted in Figure
\ref{fig:pca}.  Although there is considerable spread in these statistics between the simulations, with the exception of the HP ($b=2$) run, they occupy a region not spanned by known planetary systems, most of which contain gas giants.  Our predicted systems all have OSS $>6$, in contrast to known expoplanet systems, but this is likely an artifact of the detection bias for close-in planets.  Our systems have intermediate values of RMC (8-20) that are poorly represented by the current catalog of known multi-planet systems.

{\it Mean motion resonances:} Because of migration, two planets may
enter an MMR where the orbital periods are integer ratios.  The
resonant angle $\phi$:
\begin{equation}
\phi = pL_1-qL_2-m\bar{\omega}_1-n\Omega_1-r\bar{\omega_2}-s\Omega_2,
\end{equation} 
must librate between two values, where p, q, m, n, r, s are integers, $L$ is the mean longitude,
$\bar{\omega}$ is the argument of periapsis, and $\Omega$ is the
longitude of the ascending node.  The subscripts 1, 2 refer to the
inner and outer planet respectively.  Outside MMR,
the resonant angle will be unbounded (i.e., will circulate) \citep{2005AJ....129.1117E}.  We
searched the final billion years of the 180 primary simulation sets
for 1:1, 2:1, 3:2, 3:1, and 4:1 MMRs.  We required any MMR to last
at least 100 kyr.  None of our systems appear to have had a MMR in that interval.  Even a system that appears to be
in 1:1 commensurability (see far bottom right of Figure \ref{fig:finalconfigs}), did not
have a bounded resonant angle.\\

\begin{figure}[htbp] 
   \centering
   \includegraphics[width=8.5cm]{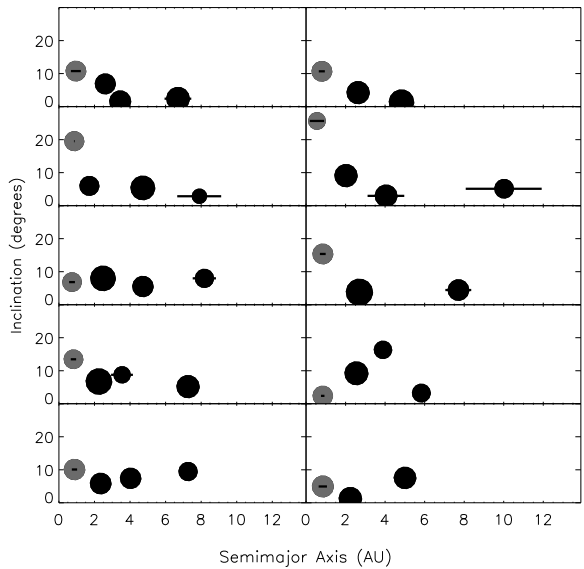} 
   \caption{Configuration of the EV3 set after 100 Myr.  The planet formed from the collision of the Earth and Venus analogs is shown in gray.  This collision occurs in all 10 simulations, forming a $2-3~M_{\earth}$ planet.  The innermost planet in the system idoes not migrate in as far inwards as in the runs where no inner mass is included, but otherwise the system is similar to the ST3 set.}
   \label{fig:EV}
\end{figure}

\begin{figure}[htbp] 
   \centering
   \includegraphics[width=8.5cm]{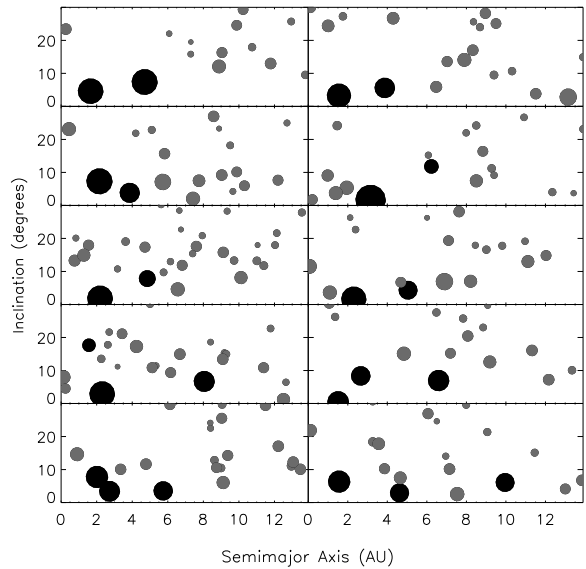} 
   \caption{Configuration of the set starting with 130 oligarchs in the inner system after 100 Myr.  Apoapsis and periapsis lines like those in Figures \ref{fig:finalconfigs} and \ref{fig:EV} are suppressed.  The surviving inner system oligarchs are shown in gray.  In most cases the inner system oligarchs were thrown onto high inclination orbits outside of the inner system or accreted by an inward migrating oligarch.  There is no obvious pattern to the distribution of the (initially) inner system oligarchs.}
   \label{fig:innerolig}
\end{figure}

{\it Inner system: } Two sets of runs (Figures \ref{fig:EV} and \ref{fig:innerolig}) in which mass (planets or
oligarchs) was placed inside the ice line show that this has little effect on the
dynamical evolution and final configuration of the outer planets.  The
innermost ice line planets were 0.1-0.2~AU further out at
100 Myr, and in the IO3 set they had accreted an average
of 0.3~$M_{\earth}$ more mass (from the inner system).  Other effects
on the outer planets were non-systematic or negligible.  However, the
effect of the outer planets on the inner system were significant.  Figure
\ref{fig:EV} shows the configuration of the EV3 system
after 100~Myr.  In all 10 runs the Earth and Venus analogs collided
after 15-70 Myr (and accreted some small bodies), forming a single
2-3 $M_{\earth}$ planet at a median semimajor axis of 0.81~AU
(Figure \ref{fig:EV}).  In each simulation, this coincides
with the time when the innermost of the outer planets migrates inside 
of 2.5~AU.  In companion runs with Earth and Venus analogs but no outer planets,
no such collision ever occurs.  After the small bodies have been
artificially removed, the configurations remain stable for at least
1~Gyr.  Figure \ref{fig:innerolig} shows the 100~Myr configuration
of systems that started with a disk of inner oligarchs rather than two
planets.  On average, 1.5~$M_\earth$ of the initial 2.2~$M_\earth$
inner disk mass has been scattered outward or accreted by the outer
oligarchs.  Amongst the 10 runs the largest surviving body in the inner system has a mass of
0.53~$M_\earth$.\\

{\it Accretion onto the central star:} There are two competing
explanations for the observed correlation between high metallicity and
the presence of giant planets.  One is that higher metallicity
augments the mass of solids in a planet-forming disk, allowing cores
to form gas giants before the gas dissipates.  The other is that rocky
material has been accreted onto the stellar photosphere during planet
formation \citep{2006PASP..118.1494G}.  Because systems without
detectable gas giants are not statistically more metal-rich than solar,
one check of our scenario is the amount of mass
that falls onto the central star.  An amount sufficient to
significantly increase the metallicity of the photosphere would
conflict with observations.  We assume that a solar mass star had a convective
region of 0.02~$M_{\sun}$ at 1 Gyr, \citep{2009ARA&A..47..481A,
1994ApJS...90..467D} and that
solids have the composition of carbonaceous chondrites (Fe is 20\% by
mass) \citep{1989AIPC..183....1G}.  In our 180 primary simulations
without mass in the inner system, the average mass accreted onto the
parent star is $0.9 \pm 0.8$~$M_{\earth}$.  For the EV3 and
IO3 sets, the average is $0.3 \pm 0.1$~$M_{\earth}$ and $3
\pm 1$~$M_{\earth}$ respectively.  The highest value in any run is
$4.8~M_{\earth}$, corresponding to $\sim 1$~$M_{\earth}$ of iron.  The
corresponding increase in [Fe/H] is no more than 0.06 dex and
more typically $\sim 0.02$ dex.  The actual metallicity enhancement is likely to
be smaller because the convective zone of solar-mass stars is much
larger at $t <$~30~Myr when much of the mass is accreted
\citep{1999ApJ...514..411F}.  Thus our scenario does not conflict with
observations.

\subsection{Sensitivity Runs}\label{sec:specialcaseresults}

The TP3 runs contain twice as many small bodies (1000) than
the other runs, but produce systems with the same mean number of planets, and
similar values of $\overline{M_{\mbox{\footnotesize tot}}}$, $\overline{\mbox{OSS}}$,
$\overline{\mbox{RMC}}$, $\overline{\mbox{AMD}}$ compared to the ST3 set (Table \ref{tab:outcomestats}). The
only major difference was that there was less variation between the
5~Gyr configurations produced by the TP3 runs.  The standard
deviation of AMD, OSS, and RMC are all lower, presumably because
random fluctuations are reduced with a larger number of small bodies.
We conclude that in most of our simulations, the between-run
variability is exaggerated due to the use of a finite number of small
bodies.

Simulations in which the small bodies were removed at 2~Gyr as opposed to 1~Gyr did not result in significantly different systems.  Most systems, including both oligarchs and small bodies, achieve a degree of stability earlier than 1 Gyr.  Runs where the initial masses, position, and Hill spacing of the oligarchs vary slightly (Table \ref{tab:initialstats}) did not produce significantly different systems.

Simulations run with higher eccentricities and inclinations for small bodies nearby the oligarchs showed slower initial mass growth than the standard set.  At 100 Myr, the HE3 set had $\sim 4~M_\earth$ less mass in oligarchs than the ST3 set.  By 250 Myr, the difference was only $\sim 2~M_\earth$.  This result supports our expectation that because the Safronov numbers are low, oligarchs ultimately accrete all planetesimals in their zone.  Inner planet migration is minimally affected by the higher eccentricities and inclinations.  At 100 Myr the innermost planet is, on average, $< 0.2$ AU further out in the HE3 set than in the ST3 set.  At 250 Myr, the difference is negligible.

\section{Prospects for Detection}\label{sec:detection}

\begin{figure}[htbp] 
   \centering
   \includegraphics[width=8.4cm]{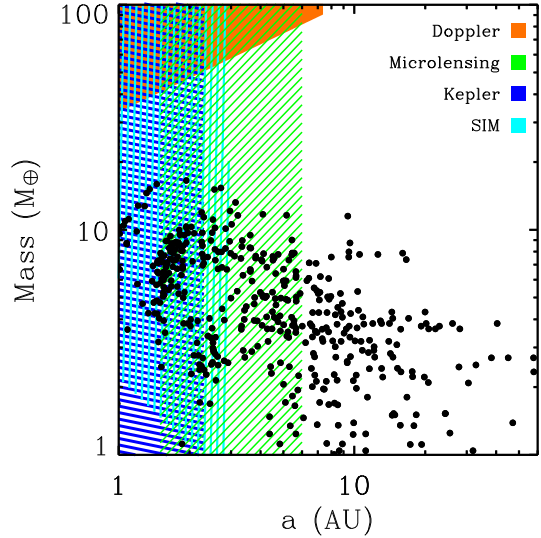} 
   \caption{Final mass and semimajor axis of all planets from the primary 6 sets and the detection domains of 4 planet-finding techniques.  The Doppler range is set by a Doppler amplitude of $K=3~$m~s$^{-1}$.  A \textit{Kepler} detection requires observation of at least 2 (rather than the usual 3) transits, and we assume a mission lifetime of 3.5 years, so planets with period $P$ $> 1.75$ yr will not be detected.  Microlensing is most sensitive in the 1.5-6~AU range, where the planet detection probability is at least $1\%$ per microlensing event.  The SIM-Lite range is set by a detection probability $> 85\%$ (see Equation (\ref{eqn.sim1}) in the text).  Although SIM might be able to detect planets with P greater than the lifetime of the mission ($\sim 5$ yr) we conservatively exclude such planets.}
   \label{fig:detectionarea}
\end{figure}

Our simulations are useful to the extent they can make testable predictions.  We investigate the prospect of detecting the 534 planets predicted by our 180 primary simulation sets.  We divide the expected detections up by initial conditions to determine which initial conditions were most accurate when detections are made.
Figure \ref{fig:detectionarea} shows their masses and semimajor axes relative
to the detection domains of Doppler radial velocity, ground-based
microlensing techniques, the NASA {\it Kepler} mission, and the proposed SIM-Lite.  Until there are substantial improvements
in sensitivity and stability \citep{2010EAS....41...27E}, Doppler is unlikely to detect any of the
predicted planets.  The other three techniques will be able to detect
at least some of these objects.

\subsection{Microlensing}
Microlensing is currently the only ground-based detection method that is sensitive to the planets predicted by our simulations.  Microlensing is most sensitive to planets with projected separations
near the Einstein radii $R_{\rm E}$ of their primaries, corresponding to
$R_{\rm E} \sim 3.5~$AU$ (M_*/M_\odot)^{1/2}$ for typical lens and source distances.
Thus for a typical primary mass in current surveys of $\sim 0.5~M_{\odot}$ \citep{2010arXiv1001.0572G}, the sensitivity of microlensing peaks for planets with semimajor axes $\sim 3~{\rm AU}$.  Current microlensing
surveys can detect planets with mass $\ga 3~M_\oplus$ with separations
within a factor of a few of this distance.  Indeed,
several of the microlensing planets detected to date have masses in
the range $3-15~M_\oplus$ and projected separations of $1-3~{\rm AU}$ \citep{2006Natur.439..437B, 2006ApJ...644L..37G, 2008ApJ...684..663B}, and thus may be analogs to our
simulated systems.  Interestingly, for the planet OGLE-2005-BLG-169Lb
with mass $\sim 13~M_\oplus$ and projected separation $\sim 2.7~{\rm
AU}$, \citet{2006ApJ...644L..37G} exclude additional Jupiter-mass planets within
the range of projected separations of $0.5-15~{\rm AU}$; indicating
that this may indeed by a system without gas giants.

We can estimate the expected number of microlensing detections one would expect, assuming that $60\%$ of stars have systems such as those we simulate.  Using the standard set, for each system we randomly choose a primary lens mass according to an event rate distribution
\begin{equation}
\frac{ {d}\Gamma}{ {d}\log{M}}
\propto M^{1/2} \frac{ {d}N}{ {d}\log{M}}
\label{eqn:eventrate}
\end{equation}
for a mass function ${
d}N/{d}\log{M} \propto M^{-\alpha+1}$.  We adopt $\alpha=0.2$ and
restrict our primary masses to the range $0.05-1~M_\odot$, with an average primary mass of $\sim 0.5~M_\odot$.  We assume the
planets in the system are coplanar and draw a random inclination $i$
for the system distributed as $\cos{i}$.  Then, for each planet, we
compute its mass ratio and projected separation, drawing a random
orbital phase for each planet, ignoring the (small)
effects of non-zero eccentricities.  We then scale the projected
separation to the Einstein radius, assuming $R_{\rm
E}=3.5~{\rm AU}(M_*/M_\odot)^{1/2}$.  Finally, we determine which of
the planets in the system are detected in each of the 13 events in the
\citet{2010arXiv1001.0572G} sample, noting instances when multiple planets are
detected.  We repeat this for all of the simulated systems and for 5000
Monte Carlo trials.  We find that \citet{2010arXiv1001.0572G} should have detected
$\sim 1.7$ planets, with an expected mean mass ratio of $\sim
10^{-4}$, and maximum mass ratio of $\sim 10^{-3.5}$.  In fact,
\citet{2010arXiv1001.0572G} found one system with mass ratio $\sim 10^{-4.1}$
(OGLE-2005-BLG-169Lb), and two systems with mass ratio $10^{-3.5}$,
consistent with our scenario.

\begin{figure}[htbp] 
   \centering
   \includegraphics[width=8.5cm]{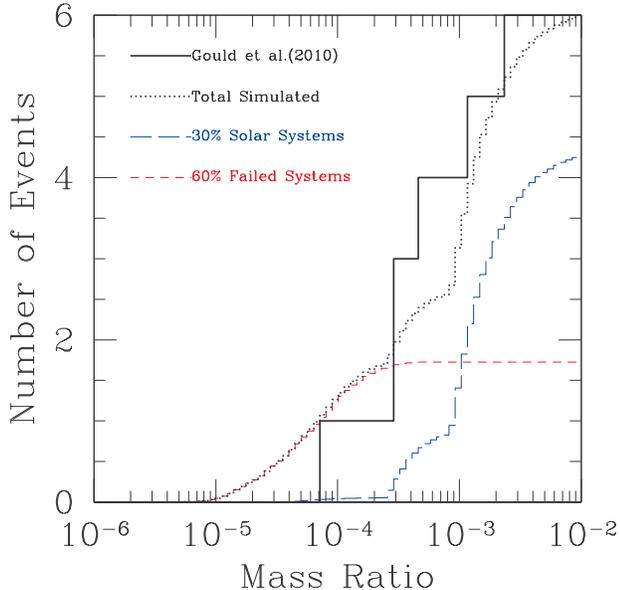} 
   \caption{The solid line shows the cumulative distribution of mass ratios for the six detected planets in the four year sample of 13 microlensing events monitored by the $\mu$FUN collaboration \citep{2010arXiv1001.0572G}.  The dotted line shows the cumulative distribution of mass ratios predicted for this sample, based on the detection efficiencies of the monitored events, and assuming a model in which 30\% of stars have four giant planets with masses and semimajor axes equal to Jupiter, Saturn, Uranus, and Neptune, and 60\% of stars have systems of planets predicted by our standard simulation set. These predictions assume a power-law distribution of primary masses, with a mean mass of $\sim 0.5 M_\odot$.}
   \label{fig:massf}
\end{figure}

Figure \ref{fig:massf} shows the observed cumulative distributions of mass ratios
from the \citet{2010arXiv1001.0572G} sample, compared to the expected
distributions for a scenario in which 60\% of stars host planets with
the properties of the standard simulation set, and 30\% host four
giant planets with the masses and semimajor axes of the solar system.
The remaining 10\% of giant planet systems host close-in planets
currently undetectable by microlensing. The number of expected
detections and the distribution of mass ratios are both broadly consistent
with the observed sample of events.  We conclude that this scenario is
consistent with all available constraints.

\begin{table*}[htdp]
\begin{scriptsize}
\begin{center}
\caption{Detection Statistics}
\label{tab:detectionstats}
\begin{tabular}{ l r r r r r r r r }
	\hline
	\hline
    Name & Kepler\footnotemark[1] & SIM\footnotemark[2] & \multicolumn{3}{c}{Microlensing\footnotemark[3]} &TTV\footnotemark[4] & Transit Prob\footnotemark[5]\\
    &&&(low-mag)&\multicolumn{2}{c}{(high-mag)} &(min)& (\%)\\
    &&&&Total&M-P Sys\footnotemark[6]&&\\
    \hline
ST&128.9&39.8&22.4&4.63&1.07&34.3& 0.30\\
LM&7.0&34.3&13.2&2.15&0.52&73.2& 0.19\\
CI&344.9&55.3&23.7&5.86&1.66&14.8& 0.47\\
MP&105.7&39.4&24.9&5.58&1.48&32.6& 0.29\\
HP&42.8&32.3&22.8&5.98&1.26&54.2& 0.21\\
LD&119.2&38.6&22.4&4.89&1.13&38.1& 0.28\\
	\hline
Average&124.8& 39.9&21.1&4.62&1.08&41.2& 0.28&\\
	\hline
\end{tabular}
\begin{minipage}{16cm}
	\footnotemark[1]{Here, a Kepler detection counts if $\ge$ 2 transits are observed over the 3.5 yr Kepler mission.  We assume $\sim 60\%$ of stars have systems similar to those in a given simulation set.}  \\  
	\footnotemark[2]{Number of detections by SIM-Lite assuming 64 target stars and that $\sim 60\%$ of stars have systems similar to those of a given simulation set.}    \\
	\footnotemark[3]{The number of microlensing detections per year assuming $\sim 60\%$ of stars have systems similar to those of a given simulation set.}\\
	\footnotemark[4]{Median transit timing variation for the innermost planets in a given simulation set.}\\
	\footnotemark[5]{Median transit probability for innermost planet in a given simulation set.}\\
	\footnotemark[6]{Number of systems with more than one planet detected in a single microlensing event.}
\end{minipage}
\end{center}
\end{scriptsize}
\end{table*}%

What are the prospects for detecting analogs to the systems we have
simulated in future microlensing surveys?  We consider the two classes of
microlensing surveys that are likely to take place over the next
10 years: alert and follow-up monitoring of high-magnification
events similar to that already being conducted, and ``next-generation'' surveys in which
thousands of low-magnification events are detected and simultaneously
monitored with the $\sim 10$ minute cadence needed to detect
Earth-mass planets using an array of $1-2$m telescopes with wide
field-of-view cameras. See \citep{2009astro2010S..85G} for further discussion of
these two channels.

For the high-magnification event channel, we follow the method
outlined above to simulate the number of expected detections, except
we assume that 20 events per year with maximum magnification
$>100$ are densely monitored during each peak.  This represents a factor of
$\sim 6$ improvement over the rate in \citet{2010arXiv1001.0572G}, which should be
realizable with the expected better prediction of high-magnification
events, increased number of alerts, and decrease in the maximum
magnification threshold from 200 to 100 \citep{2010arXiv1001.0572G}.  We adopt the
analytic detection sensitivity estimate discussed in \citep{2010arXiv1001.0572G},
assuming $\eta=0.32$ and $\xi=100$.  The results are shown in Table
\ref{tab:detectionstats}, for the six primary simulation sets.  We expect an average of 4.6 planet detections per year (for the standard set), with roughly one detection of a multiple-planet system per year.  For the HP simulation set, we find that there is a significant chance (0.17 per year) of detecting as many as four planets in the same event, whereas these probabilities are generally substantially smaller ($\le$0.03 per year) for the other simulations.  This indicates it may be possible to distinguish between the various input assumptions of the simulations using observations of multiple planet systems.

 \begin{figure}[b] 
   \centering
   \includegraphics[width=8.4cm]{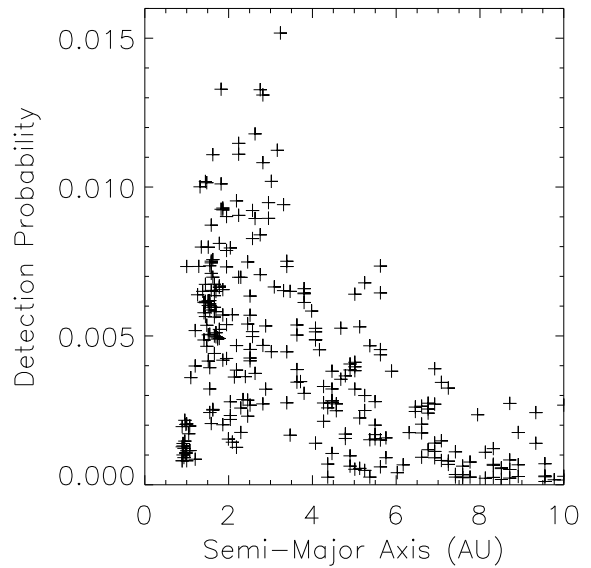} 
   \caption{Probability of detecting the planets from our 180 primary simulation sets in a next-generation, ground-based microlensing survey, consisting of three 1.6m telescopes with large FOV cameras located in Chile, South Africa, and Australia.  The higher detection probabilities near $\sim 2-3$AU are caused by their proximity to the Einstein ring and the tendency for closer in planets to have higher masses (Section \ref{sec:outcomes}).  Assuming 60\% of stars systems analogous to those in our primary simulation sets, such a next-generation ground-based microlensing survey would detect $\sim 22$ planets per year (see Table \ref{tab:detectionstats}).}
   \label{fig:microlens}
\end{figure}

For the low-magnification events detected in next-generation surveys,
we use the unpublished simulation code of Gaudi, Han, and Gould.  This
code simulates ensembles of planetary microlensing events and
estimates detection rates for a given input value of the mass and
semimajor axis of the planet.  The simulated light curves account for
the effects of weather, variable seeing, moon and sky
background, and the finite size of the source star.  We assume
parameters similar to that expected for the funded Korean Microlensing
Telescope Network next-generation microlensing survey: three
1.6m telescopes with 4 $deg^2$ cameras located in Australia, Chile, and South Africa
(C. Han, pers. communication).  The host lenses
are drawn from a model of the Galactic population of lenses and sources
that matches available constraints \citep{1995ApJ...447...53H,2003ApJ...592..172H}.  The
resulting detection probability for each of the planets in the 180
simulations is shown in Figure \ref{fig:microlens}, and the expected number of
detections per year are listed in Table \ref{tab:detectionstats}.  We predict that
next-generation surveys should detect $\sim 22$ planets in low-magnification events per year (for the standard set),
assuming that 60\% of all stars host planetary systems such as those
we simulate.  These detections are in addition to
those found in high-magnification events.  Multiple-planet systems will be rare (detection probabilities of $\la 0.1\%$) in these low-magnification events.

While ground-based surveys are relatively insensitive to the low-mass,
large semimajor axis planets we typically find in our simulated
systems, a space-based microlensing survey
\citep{2002ApJ...574..985B,2009astro2010S..18B,2010arXiv1001.3349B} would be exquisitely sensitive
to these bodies (and essentially all of the planets we find in our simulations).  In
particular, a space-based microlensing survey would detect the most
distant planets with $a\ga 15~{\rm AU}$ as isolated, short time scale
events without the signature of the host star \citep{2005ApJ...618..962H}.

\subsection{\textit{Kepler}} 
The {\it Kepler} spacecraft was successfully launched on 2009 March 6 and is continuously monitoring $\sim10^5$ F- to K-type stars with the primary objective of discovering transiting Earth-mass planets on 1~AU (1~yr period) orbits \citep{2010ApJ...713L..79K}, although the detection of many planets on shorter-period orbits is expected, e.g., \citet{2007Icar..191..453S}.  Three transits will be required to confirm a planet; hence the 3.5~yr nominal mission lifetime.  However, \citet{2008ApJ...688..616Y} point out that {\it Kepler} should detect one or two transits by planets on more distant orbits. The innermost planets in our 180 primary simulation runs have a median $a = 1.66$~AU ($P = 2.14$~yr), making it possible that two (but usually not three) transits would be observed, geometry permitting.  We calculated the expected number of such planets that {\it Kepler} will detect transiting at least twice using Equations (2) and (4) from \citet{2008ApJ...688..616Y}, assuming that 60\% of all solar-type stars have such systems, and ignoring the effect of eccentricity.  (The median eccentricity is 0.08).  We use {\it Kepler's} precision given in \cite{2010ApJ...713L.120J} and the characteristics of  {\it Kepler's} target stars from \cite{2010ApJ...713L.109B}.  The predicted number of transit detections is 129 for the standard set, and will be larger if ice lines are located closer to stars (See Table \ref{tab:detectionstats}).  Around a solar mass and radius star {\it Kepler's} detects a transit with a signal-to-noise ratio (S/N)
\begin{eqnarray}
S/R \approx 10\left( \frac{R_*}{R_\Sun} \right)^{-3/2}\left( \frac{M_*}{M_\Sun} \right)^{-1/6} \nonumber \\
\left( \frac{r_p}{R_\oplus} \right)^{2}\left( \frac{P}{3.5~yrs} \right)^{1/6}10^{-0.2(V-12)},
\end{eqnarray}
where $R_*$ and $M_*$ are the radius and mass of the star, $r_p$ is the radius of the planet, and $P$ is the period of the planet \citep{2008ApJ...688..616Y}.  The radius of a 10~$M_{\earth}$ body composed of equal parts water ice and rock/metal is predicted to be $2.3~R_{\oplus}$ \citep{2009ApJ...693..722G}.  At the median semimajor axis of our innermost planets ($a=1.66$~AU, $P=2.14$~yr) a planet orbiting a $V=12$, Sun-like star would produce a transit with a depth of 0.4~mmag and a S/N of $\sim 50$.  The depth and S/N will be larger if the planet has a thick atmosphere.

 \begin{figure}[htbp] 
   \centering
   \includegraphics[width=8.5cm]{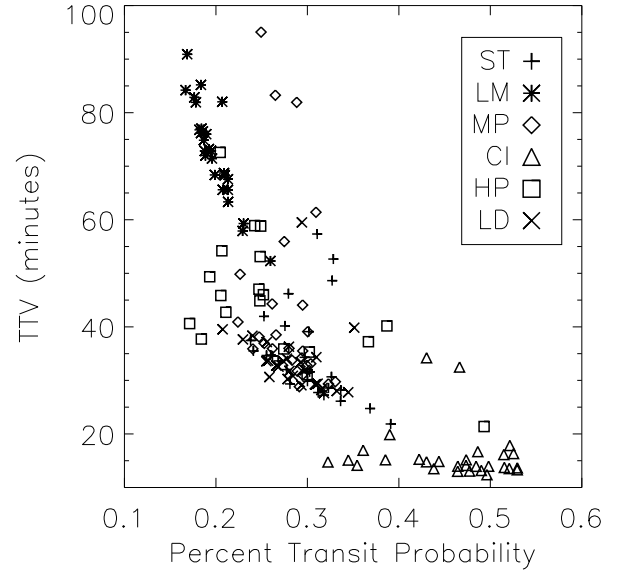} 
   \caption{Transit timing variations vs. transit probability of the innermost planets predicted by our simulations.  The TTV is the absolute difference in the time of transit center between successive transits, averaged over successive transits.  The estimated TTV precision of \textit{Kepler} ($1\sigma$) is 4 minutes in short cadence mode.  TTV scales as orbital period P, and transit probability as $P^{-2/3}$, so that TTV $\sim$ (transit probability$)^{-3/2}$.}
   \label{fig:ttv}
\end{figure}

In addition to the orbital period $P$ (the interval between transits)
it may be possible to limit the orbital eccentricity of such
transiting planets \citep{2008ApJ...678.1407F}.  To the extent that limb darkening effects can be
accounted for, the orbital eccentricity can be constrained
independently of the transit impact parameter by measuring the
duration of both the transit $T$ and the ingress and egress phases
$\tau$.  The following relation holds:
\begin{equation}
\frac{1-e^2}{\left(1+e \sin \omega \right)^2} =  \left(\frac{T\tau}{\sqrt{\delta}}\right) \left(\frac{\pi^2 G \rho_*}{3P}\right)^{2/3},
\end{equation}
where $\rho_*$ is the density of the star, and $\omega$ is the
argument of the periapsis \citep{2003ApJ...585.1038S, 2005ApJ...627.1011T}.  To the first order in eccentricity, the term on
the left is $1 - 2e \sin \omega$ and therefore only a minimum
eccentricity can be established independently of the longitude of
periapsis.  However, the distribution of $\omega$ in a population of transiting systems must be uniform over
$0-2\pi$ and thus the statistical distribution of $e$ can be
inferred.  $\rho_*$ is well
established from astrophysical theory but will not be measured
directly except in those cases where there is a second transiting
planet either on a shorter period orbit that is accessible to the Doppler
technique \citep{2007ApJ...664.1190S} or via their mutual transit timing perturbations \citep{2005Sci...307.1288H}.

The presence of an additional, outer planet of mass $M_2$ on an orbit
with semimajor axis $a_2$ can be inferred through variations in the
interval between transits \citep{2002ApJ...564.1019M,2005MNRAS.359..567A, 2005Sci...307.1288H}.
The standard deviation of the transit time due to perturbations from
the outer planet is approximately
\begin{equation}
\delta T_c \approx  \frac{P_1}{2^{3/2}\pi}\frac{a_2}{a_1}\frac{M_2}{M_*}
\end{equation}
for nearly circular orbits \citep{2005MNRAS.359..567A}.  This will be typically $3
\times 10^{-6}P_1$, or $\sim 5$~min.  If the second planet is near a
mean-motion commensurability, the variation can be 1-3 orders of magnitude
larger.  We calculated the $\delta T_c$ of the inner planet
induced by the other planets in each of the 180 systems.  These
calculations used the orbited elements computed by Mercury for 100
successive orbits at 5~Gyr.  The linear perturbations of the orbital
equations are:
\begin{equation}
\delta T_c = \delta t_0 + \frac{P}{2\pi}\delta \mu + \frac{\mu}{2\pi}\delta P,
\end{equation}
where
\begin{equation}
\delta \mu = (1 - e \cos \eta)\delta \eta - \left(\sin \eta \right) \delta e,
\end{equation}
and
\begin{eqnarray}
\delta \eta = -\frac{\cos \omega \left(1- e \cos \eta \right)^2 \delta \omega}{\left(1 + e \right) \sin \eta - e \sin \eta \cos \eta } \nonumber \\
 + \frac{\left[1 - \left(1+e\right)\cos \eta \right]\delta e}{\left(1 + e \right) \sin \eta - e \sin \eta \cos \eta},
\end{eqnarray}
where $t_0$ is the time of periapsis passage, $\mu$ is the mean anomaly, and
$\eta$ is the eccentric anomaly.  Figure \ref{fig:ttv} plots the $\delta T_c$
versus transit probability for the innermost planets in the 180 simulation
sets.  Table \ref{tab:detectionstats} has the $\overline{\delta T_c}$ and transit probability values for each simulation set.  Detection of TTV with {\it Kepler} obviously requires a
third transit, and thus an extended mission, as well as read-out in
short-cadence (59~s) mode to capture the ingress or egress (about
20~min in duration).  The S/N of the transit detection over the ingress
and egress is about 5, and the $1\sigma$ precision of the timing
(using both ingress and egress) is about 3~min.  Thus additional
observations of transits by the innermost planet by {\it Kepler}
should be sufficient to reveal the presence of outer planets like
those predicted by our simulations. Observations from the ground have
achieved $\sim0.5$ mmag precision
\citep{2009ApJ...692L.100J, 2009MNRAS.399..287S}, raising the possibility that ground-based
follow-up might also reveal such variation.

\subsection{Space Interferometry Mission (SIM-Lite)}

SIM-Lite is an astrometric interferometer mission that will achieve
sub-microarcsecond precision per ``visit'' and should be capable of
detecting Earth-mass planets in the habitable zone of nearby ($d <
30$~pc) stars \citep{2009PASP..121...41S}.  During a nominal program that consumes
40\% of a 5 yr mission lifetime, 64 target stars can each be
visited about 200 times.  The total S/N in $N$ visits is
\begin{equation}
\label{eqn.sim1}
S/N = F \times \frac{\sqrt{N}}{\sigma} \frac{a}{D}\frac{m_p}{M_*},
\end{equation}
where $m_p$ is the mass of the planet, $\sigma$ is the
measurement error in arc-seconds, $D$
is the distance in pc, and the dimensionless factor $F$ is
\begin{equation}
\label{eqn.sim2}
F = \sqrt{\frac{1}{2}\left(1 + \cos^2 i\right)\left[1-e^2\left[3- \left(\frac{2}{1+\cos^2 i}-1\right) \cos 2 \omega \right] \right] }
\end{equation}
(K. Mogren \& B. S. Gaudi, i preparation, J. Catanzarite, private communication).  
Although planets with orbital periods longer than the mission lifetime might be
detectable, we conservatively assume that this is not the case.
Equation (\ref{eqn.sim2}) is averaged over an isotropic distribution of
values for $i$ and $\omega$, and the S/N is calculated for the 64 target stars for SIM-lite provided by J. Catanzarite.  We adopt a detection criteria of S/N $> 5.8$ for a single planet \citep{2006PASP..118.1319C}.  We find that 96\% of the innermost planets will be detected.  If we assume $\sim 60\%$ of stars have systems like those in our simulations and ignore planets with periods greater than the lifetime of the mission, we predict that SIM-lite will find 44 planets like those in our simulations (see Table \ref{tab:detectionstats}).  The presence of multiple planets may make disambiguation of orbital
parameters difficult, but detection of additional planets in these systems is clearly possible.  See \cite{2006PASP..118..364F} and \cite{2008arXiv0807.4323G} for more robust analyses of this problem.

\section{Discussion}\label{sec:discussion}

\subsection{Summary}

Our simulations predict the evolution, final configuration and
detection of systems of icy planets lacking gas giants.  If, as
observations indicate, all solar-mass stars are born with disks, but
only a minority ($\sim$40\%) form giant planets, our
predictions may describe the hitherto ``invisible'' majority of outer
planetary systems.

Because planet formation is so poorly understood, and there are few
constraints on our simulations, we explore a wide range of initial
conditions.  We assume the canonical theory of planet formation
in which a small number of oligarchs grow from a disk of much
smaller planetesimals.  Motivated by the core accretion theory of
giant planet formation, we also assume that oligarchic growth was
accelerated in a region of enhanced surface density of solids beyond
the ice line, a hypothetical point in the planet-forming disk where
water ice condenses.  We do not simulate the formation of these
oligarchs, only their subsequent growth and dynamics after the gas has
disappeared from the disk.  We assume that the longer orbital periods
and lower surface density has stymied oligarch formation further out
in the disk.  The residual disk is represented by a finite number (typically
500) of small bodies.  For increased computational efficiency, we place
mass interior to the ice line in only a few realizations to
investigate its effect.

We vary the total mass in the ice line, the location of the ice line,
the initial number of oligarchs, and their spacing.  We evolve each
system for 5~Gyr; after the first Gyr the small bodies are removed.
These systems are highly chaotic in the first 10~Myr, but usually
become stable well before 1~Gyr, and removal of small bodies has no
significant effect on the oligarchs' subsequent evolution.  We
describe the evolution of these systems in terms of the time to clear
the oligarch zone of small bodies, the number of MMR
crossings, the number of oligarch ``swaps'', the migration of the
innermost oligarch, and the frequency of oligarch ejection (Table
\ref{tab:evolutionstats}).  We describe the final configuration of the
systems at 5 Gyr using three statistics: the OSS, RMC, and AMD (Table
\ref{tab:outcomestats}).  In a limited set of observations, we place
mass interior to the ice line, either pairs of Earth Venus analogs, or
smaller oligarchs, and we investigate its effect on the evolution of
the outer system, as well as its own fate.

\subsection{Major conclusions and implications}

In the vast majority (169/180) of our primary runs, and {\it
across all initial conditions we investigate}, an oligarch migrates interior to the ice line, settling to
between 25\% and 60\% of the ice line distance in about 10 Myr and growing into a planet with a 
median mass of $0.23~M_{\mbox{\footnotesize ice}}$.  We call this object the {\it innermost migrated planet}, or IMP.  In 123 of the 169 primary runs with an IMP, the IMP is the most massive
planet.  IMPs are clearly distinguishable from the other planets
(Figures \ref{fig:finalconfigs} and \ref{fig:masterMA}).  The migration
is a result of exchange of angular momentum between the IMP and the
other, exterior oligarchs: 5 of the 11 runs that did not produce an
IMP contain only a single planet at 5~Gyr.  The migration is significant because, unlike with gas giants, the planets in our simulations have low masses compared to the residual disk mass.  The existence of mass in
the inner system only slightly affects the final position and mass of
IMP, but the converse is not true (see below).  IMPs may be the visible representatives of an otherwise ``invisible''
majority: The common occurrence, relatively high mass, and small
semimajor axis of IMPs make them eminently detectable by
microlensing, transits (with {\it Kepler}), and astrometry (with
SIM-Lite), but not yet by current Doppler capabilities. 

Ground-based microlensing is currently capable of detecting planets as small as $\sim 3~M_\oplus$ at separations of 1.5-3 AU, and indeed several planets with (uncertain) masses between a few $M_\oplus$ and one or two Neptunes have been found in this distance range \citep{2006Natur.439..437B,2006ApJ...644L..37G,2008ApJ...684..663B,2010ApJ...710.1641S}.  These few detections may represent only the tip of the IMP-berg: all available constraints on the frequency of gas-giant and lower-mass planets from current radial velocity and microlensing surveys are consistent with the scenario that the minority of stars host gas giants, and that at least $\sim 60\%$ of stars host systems such as those we have simulated \citep{2006ApJ...644L..37G,2010ApJ...710.1641S}. Future microlensing surveys will provide a definitive statistical measurement or upper limit on the frequency of systems like those predicted here.  If $60\%$ of stars indeed host systems similar to those we simulate, we estimate that next-generation ground-based microlensing surveys \citep{2007arXiv0704.0767G,2008arXiv0807.4323G,2009astro2010S..85G} will detect $\sim 26$ planets per year, including a handful of multiple-planet systems. A space-based microlensing survey would be sensitive to essentially all of the planets we have simulated \citep{2002ApJ...574..985B,2009astro2010S..18B,2010arXiv1001.3349B}.

Assuming an ice-rock composition, all IMPs predicted here would
produce a transit sufficiently deep to be detected by {\it Kepler}.
83\% have periods less than the spacecraft's 3.5~yr mission.  If IMPs
are present around 60\% of solar-type stars, we predict that {\it
Kepler} will detect $\sim$129 of them with two or more transits.  Observations
of additional transits in high cadence (1~min resolution) mode in an
extended {\it Kepler} mission could reveal additional, exterior
planets through the variation of the timing of transits.  Direct
calculations show variations of 20-90~min in our predicted systems.
Finally, SIM-Lite should be able detect $96\%$ of IMPs.

The planets we predict have, statistically, very different orbital, mass, and eccentricity distributions than the  giant planets discovered to date by Doppler surveys.  The orbital eccentricities of our predicted planets are significantly lower than in known exoplanetary systems.  Our results agree qualitatively with the simulations of systems of Neptune-size  planets by \cite{2010ApJ...711..772R}.  The exceptions are, unsurprisingly, our sets
of simulations with dynamically overpacked oligarchs, whose orbital
eccentricities at 5~Gyr resemble the observed exoplanet
distribution.  {\it Kepler} observations of the duration of transit
ingress and/or egress will offer limited constraints on the
distribution of eccentricities of IMPs, if they are sufficiently
numerous.

If many more IMPs are found than predicted here, and they are closer
to their parent star, then a possible explanation suggested by our
simulations is that the ice line is often much closer than 5~AU from
the parent star.  Conversely, if the combination of microlensing, {\it
Kepler}, and SIM-Lite fail to discover a population of IMPs, one or
more assumptions in our scenario is false.  The most likely suspect
would be the assumption of a significant concentration of solids at or
immediately beyond the ice line.  This would have ramifications for
the core accretion theory of giant planet formation.

Intriguingly, inner systems of two dominant planets are not stable in
our scenario.  In all simulations with Earth and Venus analogs, the
two bodies collided, forming a single body at $\sim$ 0.8~AU.  No
contradiction with the solar system is engendered because it
contains giant planets.  Such a conglomerate would induce a
barycenter motion of 0.2~m~s$^{-1}$ (260~d period) which may be
detectable by future ultra-high precision Doppler monitoring \citep{2008PhST..130a4007P}.
On the other hand, if planet formation in the inner system has progressed only to
the giant impact (oligarch-dominated) phase by the time the IMP migrates inward, the IMP will clear most
of this mass, leaving only small ($<$ 0.3~$M_{\earth}$) bodies.

If disruption of the inner system does occur, the IMP is left as the
only detectable planet near, but exterior to, the nominal habitable zone of an Earth
``twin'' (0.95-1.37~AU) \citep{1993Icar..101..108K}.  However, it is expected that surface
temperatures will be higher on more massive planets with thicker
atmospheres, such as could be the case for the IMP.  Given that IMPs
have large quantities of water, it then follows that IMP-like planets
could be the most numerous type of habitable planet in the universe.

\subsection{Limitations of our simulations}\label{sec:limitations}

We have ignored Type I migration in our simulation.  However, the
magnitude and even the sign of Type I migration is not yet clear
\citep{2009ApJ...690L..52L, 2009ApJ...701...18M, 2009ApJ...699..824O,2010ApJ...712..198Y}.  Our results are
best seen in the context of being an end-member of a larger suite of
scenarios in which Type I migration plays a role to a varying degree, c.f. \cite{2008ApJ...673..487I}.

Planetesimals will fragment (rather than accrete) if the collisional
energy exceeds the strength of the colliders, and the production of
smaller fragments can eventually produce a collisional ``cascade''
whose ultimate product is micron-sized dust which will be swept from
the disk by radiation forces or coupling to the gas disk \citep{2008ARA&A..46..339W}.
After the gas disk disappears, the orbital eccentricities and
inclinations of planetesimals are no long damped, and they will be
excited by the oligarchs, which are growing by accretion of
planetesimals.  Thus fragmentation will compete with oligarch
accretion and may limit oligarch growth to some maximum mass.  In a
collisional cascade, most of the mass will be in the largest
planetesimals and thus it is their lifetime that will set the balance
between accretion and fragmentation.  The strength of these will be
set by gravity and will depend on on size as $s^{3/2}$ \citep{2005Icar..174..105K}, and thus the critical collision speed for
fragmentation will scale as $s^{3/4}$, e.g., 1~km~s$^{-1}$ for a
100~km body \citep{2008ApJ...673.1123L}.  The equilibrium velocities of the
planetesimals will scale with those of the oligarchs by the ratio of
mass surface densities in the respective populations
$\left(\frac{\Sigma_l}{\Sigma_s}\right)^n$ where $n \approx 0.25-0.5$
(e.g., \citet{2004ApJ...614..497G}).  Thus as planetesimals are
accreted and oligarchs grow, the collision speeds of the former will
increase until the largest planetesimals suffer destructive
collisions, after which that process competes effectively with
accretion for mass.  Unfortunately, neither the size of the largest
planetesimals (which may ultimately derive from physics such as
two-stream instabilities \citep{2007Natur.448.1022J}.  However, once $\Sigma_l \sim \Sigma_p$,
$v_p$ will become comparable to the escape speed of the oligarchs
($\sim 10$ km~s$^{-1}$) and probably well above the threshold for
fragmentation.  Thus a crude upper limit on the effect of fragmentation
is the mass of the oligarchs when the surface density of
planetesimals fall below the surface density of oligarchs.  In the standard set of simulations, this is usually after $\sim 100$~Myr.  Direct evaluation of the RMS encounter velocity of planetesimals in our standard sets show that this kinetic energy increases with time until it reaches $10\%$ of the orbital kinetic energy (i.e., $v \sim 4$~km~s$^{-1}$) in $\sim 100$ Myr.  Oligarchs have accreted $\sim 80\%$ of their mass by this time.  Orbital migration, which depends on the mass surface density of the planetesimals, and
not their mass distribution, will be affected less.

Our scenario assumes a significant concentration of mass within 1~AU
of the ice line, with a total amount sufficient to produce least one
and as many as four of the supposed 5-10 $M_{\earth}$ cores of the
outer planets in our solar system.  If the total mass in the ice line was less,
or it was less concentrated then we assume, the result would be smaller bodies, and
the amount of inward migration by the innermost object would be less.

Our systems are all orbiting solar-mass stars, whereas those surveyed by
{\it Kepler} comprise a range of stellar masses \citep{2010ApJ...713L.109B}, and the preponderance of microlensing
stars are lower mass M dwarfs ($<0.5M_{\odot}$).  Observations suggest
a correlation between stellar mass and giant planet frequency, at
least on detectable orbits \citep{2007ApJ...670..833J}, and in
line with some theoretical expectations \citep{2008ApJ...673..502K}.  At a given ice line distance, the orbital time scale is longer and the Safronov number scales inversely with stellar mass.  We
thus expect those systems around M dwarfs to develop more slowly and scattering to be
more efficient relative to accretion.  A prediction of the latter is a
higher mean orbital eccentricity among the planets than the values
found here.  However, a correlation between disk mass and stellar mass, if one exists \citep{2000prpl.conf..559N, 2008ApJ...683..304E,2009ApJ...692.1609V}, along with our finding that the mass
of the IMP approximately scales with the mass in the ice line, would
partially offset this effect.  Moreover, the ice line itself may be
closer to the star \citep{2008ApJ...673..502K}, and thus both the
orbital time scale and Safronov number will depend only weakly on
stellar mass.

Given these unresolved issues, our findings should be considered a
series of predictions of one class of planets that could be (and
perhaps is being) discovered by microlensing, {\it Kepler}, and a
future SIM-Lite mission.  If gas giant-containing systems are indeed
in the minority, then systems of icy Earth-to-Neptune-mass planets may be the
hitherto undiscovered majority of planet systems, and their innermost
members - the IMPs - could be one of the most common abodes
for life in the universe.  Assuming the continued success of microlensing surveys and 
planet-finding missions like {\it Kepler}, we shall soon know the
answer.

This research was supported in part by the National Science Foundation
through TeraGrid \citep{teragrid:2007} resources provided by Purdue
University.  Sean Raymond provided code to search for MMR.  AM and EG are supported by NSF grant AST0908419. We thank
Joe Catanzarite for assistance with the SIM-Lite detection
calculations and providing a nominal target list, Andy Gould and
Cheongho Han for permission to use unpublished simulations,
and Karen Mogren for permission to include results from
work in preparation.  We thank an anonymous reviewer for helpful comments and suggestions.


\bibliography{fullbiblio}



\end{document}